\documentclass[aps,prb,twocolumn,showpacs,floatfix,superscriptaddress]{revtex4}

\usepackage{amsmath,amssymb}
\usepackage{graphicx,color}
\usepackage{times}
\usepackage[colorlinks,citecolor=red]{hyperref}

\usepackage{natbib}

\newcommand{\bfB}{{\mathbf{B}}}
\newcommand{\bfu}{{\mathbf{u}}}
\newcommand{\bfx}{{\mathbf{x}}}
\newcommand{\bfy}{{\mathbf{y}}}

\newcommand{\bfsigma}{{\boldsymbol{\sigma}}}

\newcommand{\tk}{{\tilde{k}}}

\newcommand{\ux}{{\hat\bfx}}
\newcommand{\uy}{{\hat\bfy}}

\newcommand{\ket}[1]{\left|#1\right\rangle}
\newcommand{\Braket}[1]{\mathinner{\langle{\textstyle#1}\rangle}}
\newcommand{\acomm}[2]{\{#1,#2\}}
\newcommand{\Pde}[2][]{\frac{\partial^{#1}}{\partial{#2}^{#1}}}
\newcommand{\pde}[3][]{\frac{\partial^{#1}{#2}}{\partial{#3}^{#1}}}

\let\up\uparrow
\let\down\downarrow

\newcommand{\eqnref}[1]{Eq.~(\ref{#1})}
\newcommand{\eqnsref}[1]{Eqs.~(\ref{#1})}
\newcommand{\Eqnref}[1]{Equation~(\ref{#1})}
\newcommand{\Figref}[1]{Figure~\ref{#1}}
\newcommand{\Figsref}[1]{Figures~\ref{#1}}
\newcommand{\figref}[1]{Fig.~\ref{#1}}
\newcommand{\figsref}[1]{Figs.~\ref{#1}}
\newcommand{\secref}[1]{Sec.~\ref{#1}}
\newcommand{\Secref}[1]{Section~\ref{#1}}

\newcommand\unit[1]{\operatorname{#1}}

\begin{document}

\title{Anomalous crossed Andreev reflection in mesoscopic superconducting ring
  hosting Majorana fermions}

\author{Minchul Lee}
\affiliation{Department of Physics, College of Applied Science, Kyung Hee
  University, Yongin 446-701, Korea}

\author{Heunghwan Khim}
\affiliation{Department of Physics, Korea University, Seoul 136-701, Korea}

\author{Mahn-Soo Choi}
\affiliation{Department of Physics, Korea University, Seoul 136-701, Korea}

\begin{abstract}
  We investigate the Majorana physics and its effect on the electron transport
  in the non-topological superconductor(NS)-topological superconductor(TS)
  double junctions of a ring geometry. We find that, depending on the ratio
  between the lengths of two topologically different regions and the
  localization lengths of the Majorana fermions formed between them, two
  completely different transport mechanisms are working: perfect crossed
  Andreev reflection (CAR) for the short NS region and perfect normal Andreev
  reflection for the short TS region. The difference is explained in terms of
  the topologically distinct properties of subgap states in the NS-TS double
  junction system, which have not been revealed so far. The exotic dependence
  of the CAR process on the magnetic flux threading the ring is uncovered and
  can be used to detect the Majorana fermions.
\end{abstract}


\maketitle

\section{Introduction}
\label{sec:introduction}

Solution to the Dirac equation, \cite{Dirac28a,Dirac30a} the first quantum
theory compatible with special relativity, is complex in general, implying that
to each particle there should exist an anti-particle with same mass but
opposite charge. Theoretically, the Dirac equation can also have a real
solution. \cite{Majorana03a} The associated particle, so-called a Majorana
fermion, must then be its own anti-particle and its charge
neutral. Furthermore, Majorana fermions satisfy non-Abelian statistics, which
can be explored for topologically protected quantum
computation. \cite{Kitaev03a,DasSarma06a,Nayak08a}

Whereas Majorana fermion as an elementary particle still remains elusive with
its direct observation facing formidable technical challenges, it appears to
be far more abundant and experimentally accessible as an emergent
quasi-particle in condensed-matter systems. \cite{Wilczek09a,Franz10a}
Earlier, it was shown that unpaired Majorana fermions can exist localized at
the ends of quantum wires with certain specific conditions, \cite{Kitaev01a}
and recently several proposals have been put forward for realistic devices
based on a semiconducting nanowire with strong spin-orbit coupling and in
proximity to a
superconductor. \cite{Sau10a,Lutchyn10a,Alicea10a,Oreg10a,Alicea11a} Also
proposed are schemes to manipulate and braid the Majorana fermions to perform
quantum gates. \cite{Alicea11a,Alicea12a,Beenakker13a}

The zero-bias peak observed in recent experiments on InSb
\cite{Mourik12a,Deng12a} and InAs \cite{Das12a} nanowire strongly suggests the
existence of Majorana fermions. It may not be a decisive evidence,
\cite{Liu12a,Pientka12a,Rainis13a} though, and some other evidences are
worthwhile. One promising direction is to investigate the supercurrent
characteristics through a Josephson junction with Majorana fermions localized
at it. \cite{Kitaev01a,Jiang11a,vanHeck12a,Pientka13a}

\begin{figure}[b]
  \centering
  \includegraphics[width=2.9cm]{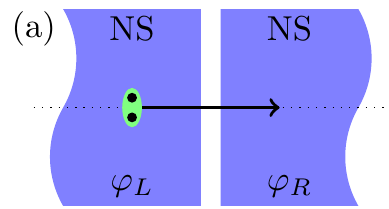}%
  \includegraphics[width=2.9cm]{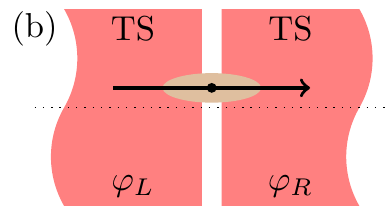}%
  \includegraphics[width=2.9cm]{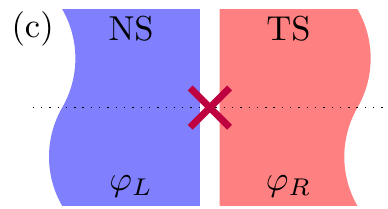}
  \caption{(Color online) (a) NS-NS junction, with ordinary Cooper-pair (double
    dot in ellipse) tunneling. (b) TS-TS junction, hosting a Dirac fermion (dot
    in ellipse) excitation at the junction. (c) NS-TS junction, hosting a
    single Majorana (cross) localized at the junction. $\varphi_L$ and
    $\varphi_R$ are the superconducting phases in the left and the right
    superconductors.}
  \label{fig:junctions}
\end{figure}

A semiconductor nanowire with strong spin-orbit coupling and in close proximity
to a superconductor turns to either topologically non-trivial superconductor
(hereafter called ``topological superconductor'' or TS) or topologically
trivial conventional superconductor (to be called ``non-topological
superconductor'' or NS). With two topologically distinct superconductors at
hand, one can consider three kinds of single Josephson junctions: NS-NS, TS-TS,
and NS-TS junctions.
The NS-NS junction is the ordinary Josephson junction in which a Cooper pair
tunnels through intermediate insulating or conducting medium [see
\figref{fig:junctions}(a)]. The supercurrent $I$ in the tunneling limit is then
a sinusoidal function of the phase difference $\delta\varphi \equiv
\varphi_L-\varphi_R$, being periodic with a period $2\pi$: $I = I_0
\sin\delta\varphi$.
%
On the other hand, the TS-TS junction hosts a single fermionic excitation
localized at the junction [see \figref{fig:junctions}(b)]. Upon the $2\pi$
change in $\delta\varphi$, a fermionic quasi-particle is transported to the
junction region and the fermion parities of the two TS regions are
reversed. Another $2\pi$ change in $\delta\varphi$ restores the fermion
parities. Thus the Josephson current exhibits $4\pi$ periodicity, unless there
is any fermion-parity-breaking process.\cite{Kitaev01a,Fu09a}
For a hybrid NS-TS junction [see \figref{fig:junctions}(c)], since both sides
have a gap and are topologically different, a single gapless state should exist
at the boundary. \cite{Kitaev01a,Alicea11a,Shen12a} The junction thus has a
single Majorana state which is pinned at the Fermi level, irrespective of the
phase difference. The supercurrent, proportional to the derivative of the
Andreev bound states with respect to the phase difference [see], should then be
zero. The vanishing supercurrent can be argued in another way: In the Majorana
state which is its own anti-particle, the amplitudes of particle and hole
excitations are the same. Further, since the particle and hole are at the same
(Fermi) energy level, their group velocities have the same magnitude. Hence,
their contributions to the current should cancel out each other exactly, and no
current flows through the hybrid NS-TS junction.

\begin{figure}[t]
  \centering
  \includegraphics[width=6cm]{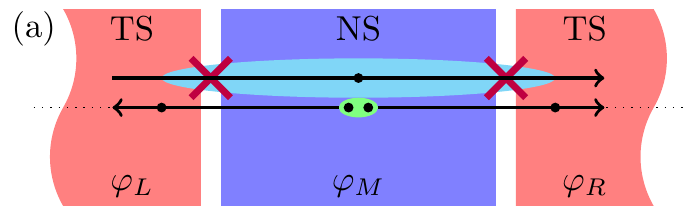}\\
  \includegraphics[width=6cm]{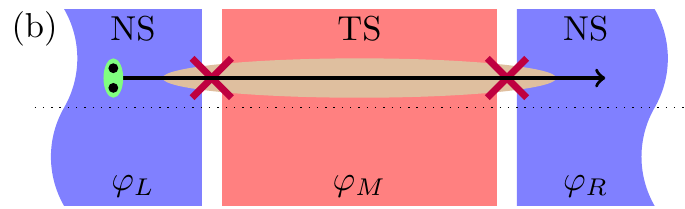}
  \caption{(Color online) (a) TS-NS-TS double junction and (b) NS-TS-NS double
    junction. $\varphi_L$, $\varphi_M$, and $\varphi_R$ are the superconducting
    phases in the left, middle, and the right superconductors.}
  \label{fig:doublejunctions}
\end{figure}

Recently, Jiang et al.\cite{Jiang11a} proposed a way to induce a supercurrent
through the hybrid NS-TS junction by making a TS-NS-TS double junction. In
their setup, the middle NS is short enough that the overlap between the two
Majorana states localized at both ends of the NS segment is finite. The overlap
couples the two Majorana states so that their energies are lifted from the
Fermi level and the vanishing current condition is no longer valid. They
predicted two different mechanisms of electron tunneling [see
\figref{fig:doublejunctions}(a)]: First, a single electron can tunnel from one
TS to the other TS like in the TS-TS junction. Secondly, a Cooper pair in the
middle NS is split, and each of two electrons from the Cooper pair tunnels into
either left or right TS. The Josephson junction energy from the two tunneling
mechanisms have different dependence on the superconducting phases: the
Josephson energy from the former process follows that of the TS-TS junction
\begin{align}
  \label{eq:EMterm}
  E_M \cos\frac{\varphi_L-\varphi_R}{2},
\end{align}
and the energy due to the Cooper pair splitting is given by
\begin{align}
  \label{eq:EZterm}
  E_Z \cos\left(\frac{\varphi_L+\varphi_R}{2}-\varphi_M\right).
\end{align}
It was proposed to measure unusual Shapiro steps in a non-local ac
current in order to detect the latter tunneling mechanism.

What about a NS-TS-NS double junction [see \figref{fig:doublejunctions}(b)]
with a short TS segment in the middle?  Interestingly, even though it is
seemingly a counter part of the TS-NS-TS double junction discussed in
Ref.~\onlinecite{Jiang11a}, its Majorana physics and associated supercurrent
characteristics are quite different.
As we will show in detail with numerically exact calculations (see
\Secref{sec:results}) and perturbation theory (see \Secref{sec:perturbation}),
the main difference is that the energy splitting due to the overlap over the TS
segment of the two Majorana states is independent of the phase difference and
does not carry supercurrent. Putting another way, NS-TS-NS and TS-NS-TS double
junctions have topologically different characteristics:
Since NS preserves the fermion parity, it cannot accept a single electron, and
the two transport mechanisms working in the TS-NS-TS double junction cannot
take place.
In principle, the overlap between the Majorana fermions opens a fermionic
channel through the short TS so that the Cooper pair in the NS regions can
tunnel through it via virtual processes as depicted in
\figref{fig:doublejunctions}(b). Hence, the Josephson energy in the NS-TS-NS
double junction will behave like
\begin{align}
  \label{eq:ECterm}
  E_C \left[\cos(\varphi_L-\varphi_M) + \cos(\varphi_M-\varphi_R)\right]
\end{align}
in a symmetric double junction. Since the Cooper pair tunneling demands the
cotunneling processes, the magnitude of $E_C$ would be significantly smaller
than those of $E_M$ and $E_Z$ in the counterpart setup.

\begin{figure}[t]
  \centering
  \includegraphics[width=7cm]{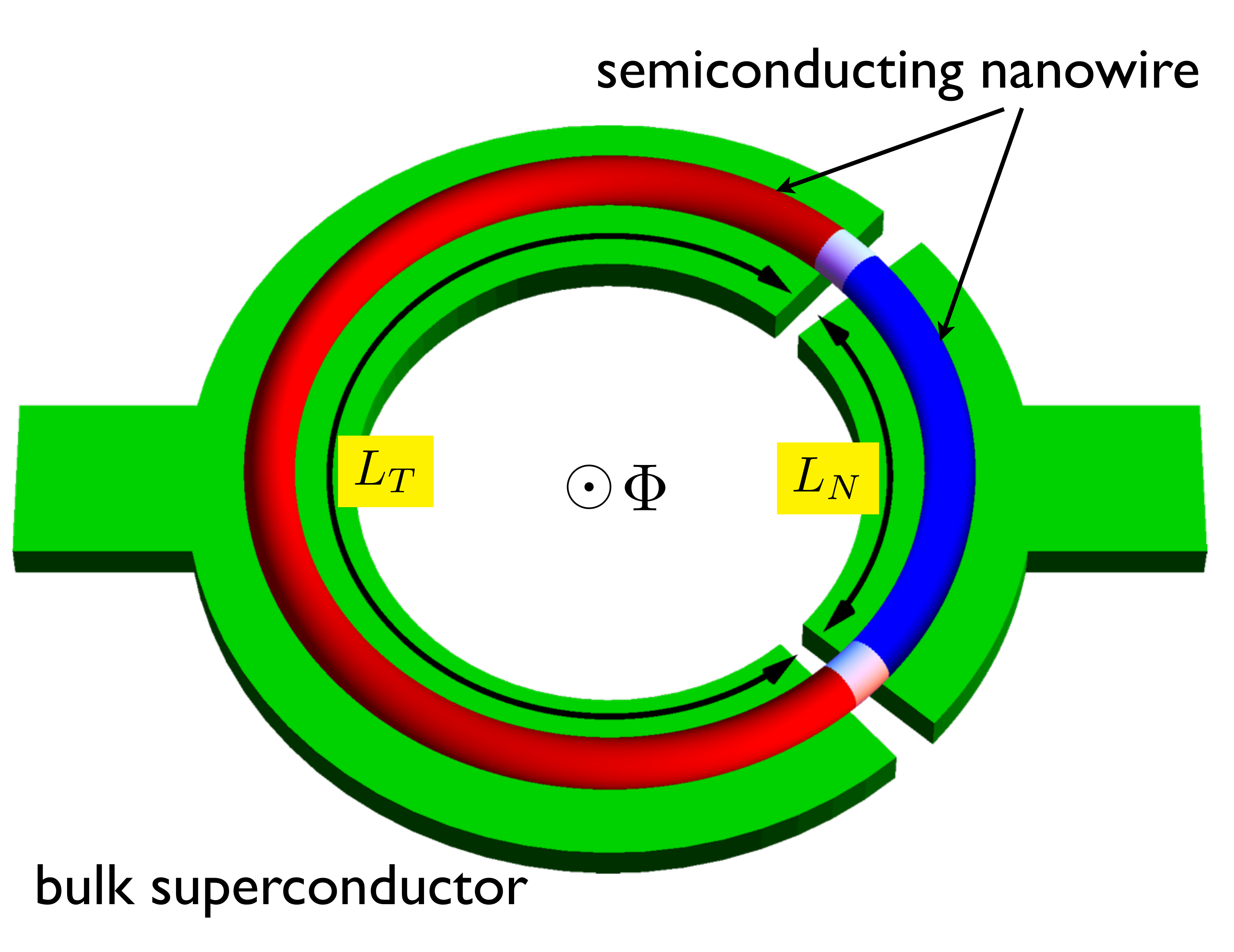}
  \caption{(Color online) Illustration of a ring made of a semiconducting
    quantum wire in proximity to two $s$-wave superconductors. The underlying
    superconductors are interrupted by insulators so that no current can flow
    directly between bulk superconductors. Different gate voltages on two
    segments of the quantum wire define two topologically different parts whose
    length are $L_T$ and $L_N$, respectively.  The external magnetic field
    $\bfB$ pierces the ring inducing the Zeeman splitting and the magnetic flux
    in the wire.}
  \label{fig:model}
\end{figure}

In this paper we investigate the Majorana physics and the corresponding
electron transport in the TS-NS double junctions of a closed ring geometry as
shown in \figref{fig:model}. This setup contains both TS-NS-TS and NS-TS-NS
double junction, allowing us to study them on an equal footing.
The Aharonov-Bohm phase from the threading magnetic flux and the phase
difference between bulk superconductors that induce the $p$-wave
superconductivity in the nanowire control the relative phases across the junctions between NS and TS.
We find that the supercurrent characteristics through the ring strongly depends
on the relative ratio between the segment lengths and the localization lengths
of the Majorana states. For short (compared with the localization lengths of
the Majorana states) NS and long TS segments (see \Secref{sec:shortNS}), the
supercurrent originates solely from the crossed Andreev reflection (CAR),
exhibiting an unusual dependence on the magnetic flux.
For short TS and long NS segments (see \Secref{sec:shortTS}), on the contrary,
the normal Andreev reflection (NAR) determines the supercurrent, whose sign can
be oscillatory with the TS segment length. The difference in the supercurrent
features of the two extreme cases is explained in terms of topological
properties in the subgap states (see \Secref{sec:topology}).
The representative characteristics in the above two extreme cases compete with
each other and show rich effects, which we study by varying the lengths of NS
and TS segment (see \Secref{sec:length} and \ref{sec:smallRings}).
Finally, we discuss the applicability of our setup to detect the existence of
Majorana states and its physics.

The paper is organized as follows: In \Secref{sec:model} we describe our system
of $p$-wave superconductor double junctions of the ring geometry and the method
to obtain the subgap states and the corresponding supercurrent. In this
section, we also discuss the elementary features of the subgap states and
associated supercurrent of topological origin, which will provide the physical
interpretations for the results to be presented in
\Secref{sec:results}. \Secref{sec:results} presents and discusses the numerical
results leaving the perturbative calculations in
\Secref{sec:perturbation}. \Secref{sec:conclusion} concludes the paper.

\section{Model and Method}
\label{sec:model}

\subsection{$p$-wave Superconductor Junctions}

We consider a narrow semiconductor ring in proximity to two spatially separated
$s$-wave superconductors as shown in \figref{fig:model}.
The semiconductor wire forms a ring geometry with radius $R$ and circumference
$L = 2\pi R$.\footnote{Even though we assume a circular ring for convenience,
  it need not be a perfect circle as long as the nanowire forms a closed
  loop. Essential physics does not depend on the specific
  geometry. Experimental realization may prefer a closed loop with straight
  semiconductor segments (instead of curved segments) in favor of easier
  layering of nanowires and superconductors.}
The underlying bulk superconductors are attached to
superconducting electrodes so that the supercurrent through the ring can be
measured.  In the ring part, two (lower and upper) junctions at $x = x_a = 0$
and $x = x_b = L_N$ are introduced by inserting insulating regions between
superconductors. Here we assume that the insulators are thick enough so that no
current can flow directly between superconductors.
In the presence of the magnetic flux $\Phi$ threading the ring, the phase of
the superconducting order parameter, $\varphi(x)$, depends on the position and
in a proper gauge is given by $\varphi(x) = 4\pi{f}x/L$, where $f \equiv
\Phi/\Phi_0$ is the dimensionless magnetic flux and $\Phi_0 \equiv h/e$ is the
flux quantum for a \emph{single} electron.
Apart from the magnetic-flux contribution, an additional phase difference is
generated between two superconductors when a bias current is applied across the
ring.
Putting them all together, the overall order-parameter phase takes the form
\begin{align}
  \varphi(x) = \varphi_0(x) + 4\pi f \frac{x}{L}
\end{align}
with
\begin{align}
  \varphi_0(x)
  =
  \begin{cases}
    \varphi_N & (x_a<x<x_b) \\
    \varphi_T & (x_b<x<L) \,.
  \end{cases}
\end{align}
Via the proximity effect, the bulk superconductors induce an $s$-wave
superconductivity on the semiconducting nanowire, on which the order parameter
is given by $\Delta(x)=\Delta_0 e^{i\varphi(x)}$.

\begin{widetext}
  Assuming that the semiconducting nanowire is narrow enough that only the
  lowest transverse mode is involved, the Hamiltonian of the
  superconductivity-induced wire then reads
  \begin{align}
    \label{eq:H}
    H
    =
    \oint dx
    \left\{
      \begin{bmatrix}
        \psi_\up^\dag(x) & \psi_\down^\dag(x)
      \end{bmatrix}
      \left(
        \frac{\Pi_x^2}{2m} - \mu_F(x) + \frac{V_Z}{2} \sigma_z
        +
        \frac{\alpha}{\hbar} \frac{\acomm{\bfsigma\cdot\bfu(x)}{\Pi_x}}{2}
      \right)
      \begin{bmatrix}
        \psi_\up(x) \\ \psi_\down(x)
      \end{bmatrix}
      + \Delta(x) \psi_\up^\dag(x) \psi_\down^\dag(x) + (h.c.)
    \right\}
  \end{align}
  with $\Pi_x = p_x - 2\pi\hbar{f}/L$.  The field operator $\psi_s(x)$
  describes the electronic degrees of freedom in the lowest transverse mode
  with spin $s = \up,\down$ and effective mass $m$ ($m\approx 0.015m_e$ for
  InSb \cite{Mourik12a,Deng12a} and $m\approx 0.03m_e$ for InAs \cite{Das12a}).
  One of the key ingredient for effective $p$-wave superconductivity is the
  strong Rashba spin-orbit coupling, which is specified by the parameter
  $\alpha$ ($\alpha\approx 0.2\unit{eV{\cdot}\AA}$ for InSb
  \cite{Mourik12a,Deng12a} and InAs \cite{Das12a}) or equivalently by the
  spin-orbit length $\ell_{so}\equiv\hbar^2/m\alpha$ ($\ell_{so}\approx
  200\unit{nm}$ for InSb \cite{Mourik12a,Deng12a} and $\ell_{so}\approx
  127\unit{nm}$ for InAs \cite{Das12a}). The Rashba-induced effective magnetic
  field is perpendicular to the wire direction and hence varies along the wire,
  and $\bfu(x) = \ux\cos\phi(x) + \uy\sin\phi(x)$ is the unit vector parallel
  to the Rashba field at the position $x$. The inner curly brackets denote the
  anticommutator, and guarantees the hermiticity of $H$ in the presence of
  position-dependent Rashba field $\bfu(x)$.
  The other ingredient is the Zeeman field $V_Z$ perpendicular to the Rashba
  field, which is applied perpendicular to the ring plane. The applied magnetic
  field should induce a finite spin splitting but be still weak enough not to
  break the superconductivity ($B \sim 100\unit{mT}$ \cite{Mourik12a,Deng12a}).
  $\mu_F(x)$ is the position-dependent chemical potential: with $\mu_F(x) =
  \mu_N$ for $x_a < x < x_b$ and $\mu_T$ for $x_b < x < L$. As will be
  discussed below, the topological state of each region is controlled by
  locally tuning the chemical potential.

\end{widetext}

The model, \eqnref{eq:H} for a uniform wire (closed or open) is exactly
solvable via the Bogoliubov-de Gennes (BdG) transformation in the chiral basis
diagonalizing the single-particle part of the Hamiltonian. \cite{Shen12a} The
two channels with chirality $\zeta=\pm$ are completely decoupled and a finite
$p$-wave pairing potential between electrons with same spin in each channel is
induced, whose order parameter is proportional to $\alpha \Delta_0/V_Z$ in the
small momentum limit. It illustrates that the Rashba spin-orbit coupling, the
Zeeman splitting, and the $s$-wave superconductivity combines together to form
three indispensable ingredients to implement $p$-wave superconductor.
Even though both channels exhibit the $p$-wave superconductivity, one of them
(say $\zeta=+$) has a finite excitation gap between particle and hole bands
remains finite at any value of momentum $k$, irrespective of the strength of
the system parameters. On the other hand, the gap for the other channel ($\zeta
= -$) closes when the parameters are properly tuned. Hence, near the quantum
phase transition point, only the $\zeta=-$ channel is relevant and one can
project out the other channel by focusing on the low-energy physics.

Since our system is piece-wise uniform, we adopt the same projection on to the
$\zeta=-$ channel to obtain an effective $p$-wave superconducting wire of
spinless fermions:
\begin{multline}
  \label{eq:Heff}
  H_\mathrm{eff}
  =
  \oint dx
  \Bigg[
  \psi^\dag(x)
  \left(\frac{\Pi_x^2}{2m_\mathrm{eff}} - \mu_\mathrm{eff}(x)\right)
  \psi(x)
  \\{}
  + \frac{\Delta_\mathrm{eff}(x)}{2} \psi^\dag(x)\partial_x\psi^\dag(x) + (h.c.)
  \Bigg]
\end{multline}
with the effective mass $m_\mathrm{eff} = (1/m - \alpha^2/\hbar^2V_Z)^{-1}$ and
the effective chemical potential $\mu_\mathrm{eff}(x) = \mu_F(x) + V_Z -
\Delta_0^2/2V_Z$. The induced $p$-wave order parameter
\begin{align}
  \Delta_\mathrm{eff}(x)
  = \frac{\alpha\Delta_0}{\hbar V_Z} i e^{i(\varphi(x) + \phi(x))}
\end{align}
has two contributions to its phase: $\varphi(x)$ inherited from the phase of
the order parameter of the bulk superconductors, and $\phi(x)$ from the
position-dependent direction $\bfu(x)$ of the Rashba field.
The corresponding BdG equation has the form
\begin{align}
  \label{eq:BdGeq}
  i\hbar\Pde{t}
  \begin{bmatrix}
    \psi(x) \\ \psi^\dag(x)
  \end{bmatrix}
  =
  H^\mathrm{BdG}_\mathrm{eff}
  \begin{bmatrix}
    \psi(x) \\ \psi^\dag(x)
  \end{bmatrix}
\end{align}
with
\begin{align}
  \label{eq:Hbdg}
  H^\mathrm{BdG}_\mathrm{eff}
  =
  \begin{bmatrix}
    \frac{(p_x - 2\pi\hbar{f}/L)^2}{2m_\mathrm{eff}} - \mu_\mathrm{eff}(x)
    & \frac12 \acomm{\Delta_\mathrm{eff}(x)}{p_x} \\
    \frac12 \acomm{\Delta_\mathrm{eff}^*(x)}{p_x}
    & - \frac{(p_x + 2\pi\hbar{f}/L)^2}{2m_\mathrm{eff}} + \mu_\mathrm{eff}(x)
  \end{bmatrix}.
\end{align}

Below we solve the effective model, \eqnsref{eq:Heff}, (\ref{eq:BdGeq}), and
(\ref{eq:Hbdg}) by first seeking the solution for each uniform wire segment and
then matching the solutions across the junctions.

\subsection{Bulk States in a Uniform Segment}
\label{sec:bulk}

\begin{figure}[t]
  \centering
  \includegraphics[width=4.2cm]{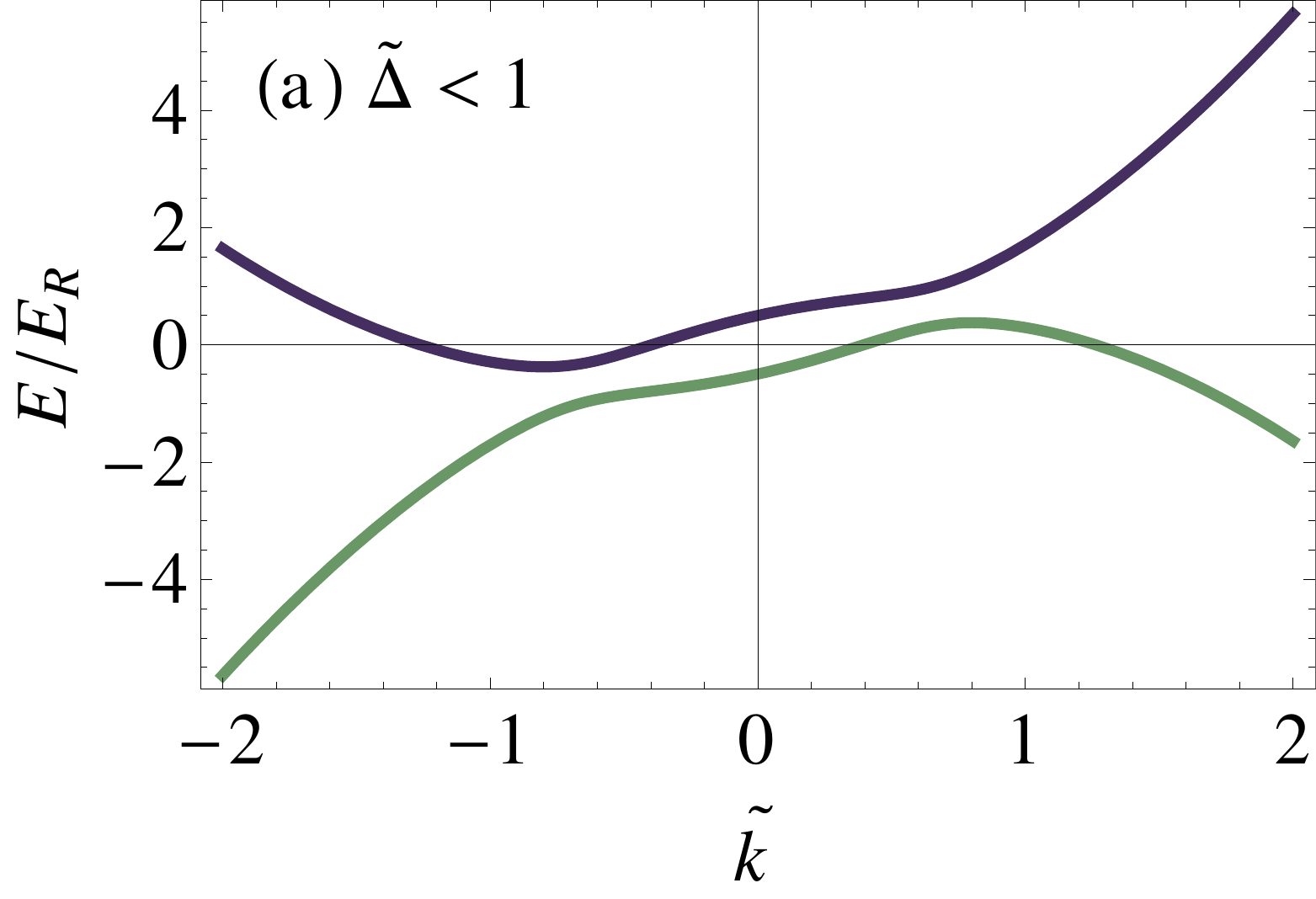}%
  \includegraphics[width=4.2cm]{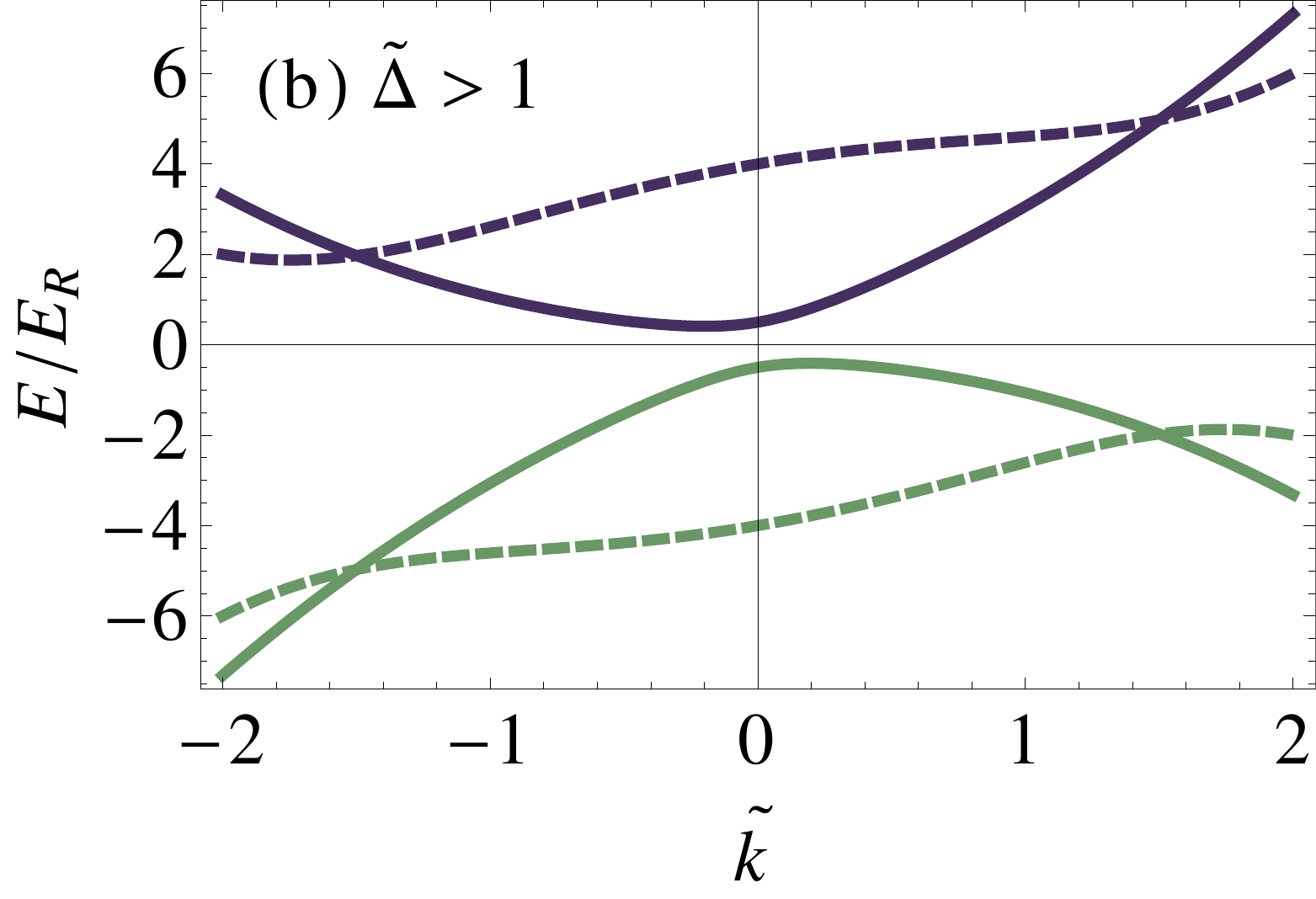}
  \caption{(Color online) Bulk spectrum for the uniform chemical potential and
    superconducting phase, \eqnref{eq:bulkspectrum} for $p$-wave
    superconducting wire of ring geometry for (a) $\tilde\Delta < 1$ and (b)
    $\tilde\Delta > 1$. The case (a) is achievable only for very small rings
    and we focus on the case (b) in this work. In figure (b), the solid and
    dashed lines correspond to the bulk spectrum in the topological phase with
    $\tilde\mu \ll D$ and $\tilde\mu \gg D$, respectively. For the definition
    of $D$, see the text.}
  \label{fig:bulkspectrum}
\end{figure}

For a uniform wire segment with $\mu_\mathrm{eff}$ and $\Delta_\mathrm{eff}$
constant, the bulk spectrum for an particle-like ($E_+$) and hole-like ($E_-$)
excitation are given by
\begin{align}
  \label{eq:bulkspectrum}
  E_\pm(\tk)
  = E_R \left(\tk \pm \sqrt{(\tk^2 - \tilde\mu)^2 + \tilde\Delta^2 \tk^2}\right),
\end{align}
with
\begin{multline}
  \tilde{k} \equiv k R \,,\quad
  E_R \equiv \frac{\hbar^2}{2m_\mathrm{eff}R^2} \,,\\
  \tilde\mu
  \equiv \frac{\mu_\mathrm{eff}}{E_R} - \frac14 \,,\quad
  \tilde\Delta
  \equiv \frac{\hbar|\Delta_\mathrm{eff}|}{E_RR}
  = \frac{2\Delta_0}{V_Z}\frac{R}{\ell_{so}}
\end{multline}
The spectrum becomes gapless for $\tilde\mu = 0$, at which occurs the
topological phase transition between a topological phase (T) with $\tilde\mu >
0$ and a non-topological phase (N) with $\tilde\mu < 0$.

The spectrum is asymmetric with respect to $\tk \to -\tk$; see
\figref{fig:bulkspectrum}. The variation of the Rashba field direction along
the curved wire invokes the precession of electron spin, and the resulting
Berry phase leads to a finite $z$ component in the spin polarization axis which
is exactly opposite for clockwise $(\tk<0)$ and counterclockwise $(\tk>0)$
movers. Adding the Zeeman field, therefore, makes the magnitude of the $z$
component different for two opposite movers, introducing asymmetry between
them. \cite{Splettstoesser03a} As a result, the gap between the particle and
hole bands is indirect. In particular, for small $\tilde\Delta < 1$, the system
is metallic over the whole range of energy; see \figref{fig:bulkspectrum}(a).
However, such an asymmetry effect is pronounced only for very small ring ($R\ll
\ell_{so}V_Z/2\Delta_0\approx 200\unit{nm}$). In our study, we therefore focus
on the case with $\tilde\Delta > 1$, where the gap $E_\mathrm{gap}$ is finite
and almost direct.

The bulk eigenstates corresponding to the spectrum, \eqnref{eq:bulkspectrum}
are
\begin{subequations}
  \label{eq:bulkeigenstate}
  \begin{align}
    \chi_{k,+}(x)
    & =
    e^{ikx}
    \begin{bmatrix}
      e^{+i(\varphi/2+(f+1/2)x/R)} \cos\frac{\vartheta_k}{2}
      \\
      e^{-i(\varphi/2+(f+1/2)x/R)} \sin\frac{\vartheta_k}{2}
    \end{bmatrix}
    \\
    \chi_{k,-}(x)
    & =
    e^{ikx}
    \begin{bmatrix}
      - e^{+i(\varphi/2+(f+1/2)x/R)} \sin\frac{\vartheta_k}{2}
      \\
      e^{-i(\varphi/2+(f+1/2)x/R)} \cos\frac{\vartheta_k}{2}
    \end{bmatrix}
  \end{align}
\end{subequations}
with the angle $\vartheta_k$ defined by $\tan\vartheta_k = \tilde\Delta\tk/(\tk^2 - \tilde\mu)$.
The phase $\pm x/2R$ in the exponents originates from the variation of the
Rashba field direction, $\phi(x)$, resulting in the Berry's phase $\pi$ for one
cycle along the ring. As can be seen from \eqnref{eq:bulkeigenstate}, this
Rashba phase always appears together with the magnetic flux $f$ in the form of
$f+1/2$. In other words, the actual effect of the Rashba phase is to apply an
additional half flux quantum $\Phi_0/2$ through the ring.

\subsection{Topological Property of Subgap States at Junctions}
\label{sec:topology}

By applying a non-uniform chemical potential along the ring as given by
\begin{align}
  \tilde\mu_\mathrm{eff}(x)
  =
  \begin{cases}
    \tilde\mu_N < 0 & (x_a < x < x_b)
    \\
    \tilde\mu_T > 0 & (x_b < x < L)
  \end{cases}\ ,
\end{align}
the two segments become topologically different superconductors, and a
localized Majorana state is formed at each interface $x =
x_{a,b}$. \cite{Kitaev01a,Alicea11a} Before obtaining exact subgap states in
the non-uniform configuration (see \secref{sec:closedring}), we examine the
topological structure of the Majorana states localized at the junctions. Here
we focus on the case of isolated Majorana states and disregard the interaction
between them. For further simplicity, we turn off the magnetic flux
($f=0$). Without the interaction, the energy of the Majorana state is zero, and
by seeking zero-energy solution in \eqnsref{eq:bulkspectrum} and
(\ref{eq:bulkeigenstate}), we obtain four complex wave vectors
\begin{align}
  \label{eq:k:zeroenergy}
  k^\ell_{\eta\nu} = \eta [(-1)^\nu k_r + i/\lambda_{\ell\nu}]
\end{align}
and the corresponding wave functions
\begin{align}
  \label{eq:wavefunction:zeroenergy}
  \chi^\ell_{\eta,\nu}(x)
  =
  e^{ik^\ell_{\eta\nu}x}
  \begin{bmatrix}
    e^{+i(\varphi_\ell/2 + x/2R)} e^{+i\gamma^\ell_{\eta\nu}/2}
    \\
    e^{-i(\varphi_\ell/2 + x/2R)} e^{-i\gamma^\ell_{\eta\nu}/2}
  \end{bmatrix}
\end{align}
for each region $\ell = T,N$. Here the index $\eta=\pm$ denotes the decay
direction of the wave function tail ($\eta=+(-)$ state decays in the positive
(negative) direction) and $\nu=1,2$ distinguishes different Majorana modes with
different localization lengths ($\lambda_{\ell1}\geq\lambda_{\ell2}$).
The wave vectors $k^\ell_{\eta\nu}$ are complex and their imaginary parts
determine the localization length of the Majorana states. In the NS region
($\tilde\mu_N<0$), all wave vectors are pure imaginary ($k_r=0$) and the
localization lengths are given by
\begin{align}
  \lambda_{N\nu}
  = \frac{R}{\sqrt{D+|\tilde\mu_N|}+(-1)^\nu\sqrt{D}}
\end{align}
with $D \equiv (\tilde\Delta^2-1)/4$.  In the TS region ($\tilde\mu_T>0$), the
real part is given by
\begin{align}
  \label{eq:kr}
  k_r =
  \begin{cases}
    0 & (0<\tilde\mu_T<D)
    \\[3ex]
    \displaystyle\frac{\sqrt{\tilde\mu_T-D}}{R} & (\tilde\mu_T>D)
  \end{cases}
\end{align}
and the localization length by
\begin{align}
  \lambda_{T\nu}
  =
  \begin{cases}
    \displaystyle
    \frac{R}{\sqrt{D}+(-1)^\nu\sqrt{D-\tilde\mu_T}} & (0<\tilde\mu_T<D)
    \\[3ex]
    \displaystyle\frac{R}{\sqrt{D}} & (\tilde\mu_T>D)
  \end{cases}\ .
\end{align}
The relative phase difference $\gamma^N_{\eta\nu} = \eta(-1)^\nu\gamma$ in the
NS region depends on both $\eta$ and $\nu$, where the angle $\gamma$ has been
defined by
\begin{align}
  \label{eq:gamma}
  e^{i\gamma}
  \equiv
  - \frac{1}{\tilde\Delta} + i \sqrt{1 - \frac{1}{\tilde\Delta^2}},
\end{align}
but $\gamma^T_{\eta\nu} = \eta\gamma$ in the TS region does not depend on
$\nu$. This difference leads to intriguing topological properties as we discuss
below.
The wave functions $\Psi_i(x)$ for Majorana states localized at $x=x_i$
($i=a,b$) are then given by linear superpositions of the eigenstates,
\eqnref{eq:wavefunction:zeroenergy}; refer their explicit forms to
\eqnref{eq:Majoranawavefunction}. The coefficients for eigenstates are
determined by the matching condition at each junction:
\begin{align}
  \label{eq:bc}
  \Psi_i(x_i^+) = \Psi_i(x_i^-) \,,\quad
  v_x\Psi_i(x_i^+) = v_x\Psi_i(x_i^-)
\end{align}
where $v_x$ is the velocity operator along the wire
\begin{align}
  \label{eq:v}
  v_x
  =
  \begin{bmatrix}
    -iR \partial_x - f & \frac{\tilde\Delta}{2} e^{+i(\varphi(x)+(2f+1)\frac{x}{R})}
    \\
    \frac{\tilde\Delta}{2} e^{-i(\varphi(x)+(2f+1)\frac{x}{R})} & iR \partial_x - f
  \end{bmatrix}.
\end{align}
Note that $x_a^- = L$.

\begin{figure}[t]
  \centering
  \includegraphics[width=8cm]{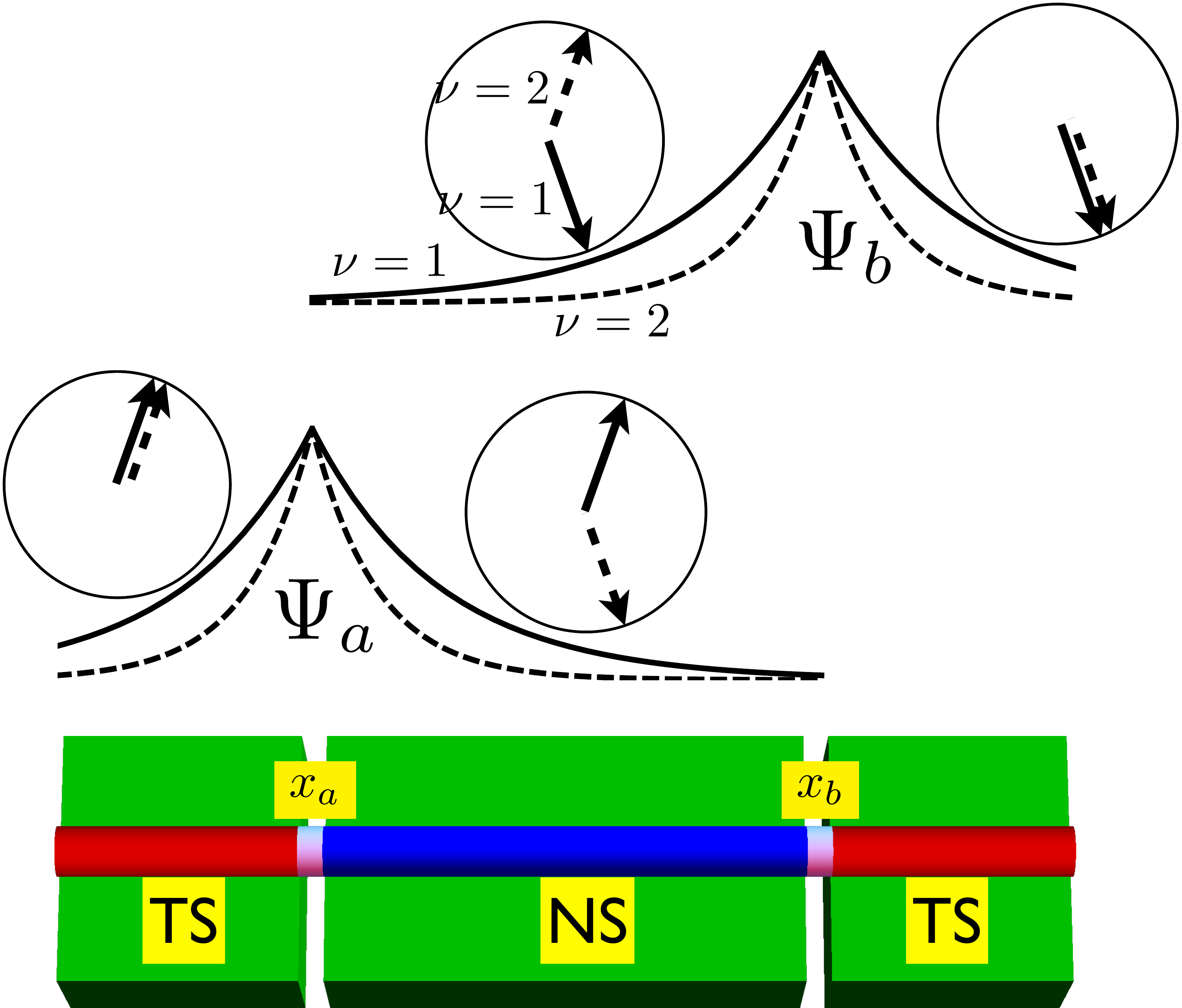}
  \caption{(Color online) A schematic representation of the subgap states,
    $\Psi_a$ and $\Psi_b$, at the TS-NS ($x_a$) and NS-TS ($x_b$) junction
    interface, respectively. The curves depict the spatial distributions of the
    wave functions, and the arrows in circles the pseudo-spin polarizations in
    the particle-hole basis. The solid and dotted curves/arrows correspond to
    $\nu=1$ and $2$ Majorana mode, respectively. Here no superconducting phase
    difference ($\delta\varphi=0$) is applied for simplicity. Note that for
    finite length of the NS segment, the wave function $\Psi_a$ should be
    matched with $\Psi_b$ properly at $x=x_b$, and $\Psi_b$ with $\Psi_a$ at
    $x=x_a$. When matching, $\nu=1$ mode undergoes a pseudo-spin rotation by
    angle $2\gamma$, while $\nu=2$ mode does not.}
  \label{fig:topology}
\end{figure}


In order to clarify the topological difference between the wave functions,
\eqnref{eq:wavefunction:zeroenergy} of the NS and TS regions, we regard the
wave functions as spinors in the pseudo-spin up ($\up$) and down ($\down$)
basis in the particle-hole (or so-called Nambu) space, and examine their
pseudo-spin polarization directions.
\Figref{fig:topology} shows the pseudo-spin polarization of the eigenstates
localized at the TS-NS and NS-TS junctions for $\delta\varphi=0$. We see the
clear difference between relative pseudo-spin polarizations in two
topologically different regions. In the NS region, the two evanescent modes
($\nu=1,2$) localized at the same end form an angle $2\gamma$ ($\pi/2 < \gamma
< \pi$), while they are parallel to each other in the TS region: For an
infinite-curvature ring $(\tilde\Delta\gg1)$, $\gamma \approx \pi/2$ so the two
modes are polarized in the opposite direction.
The phase difference $\delta\varphi = \varphi_N - \varphi_T$ leads to the
misalignment between the polarization axes for the two regions. \footnote{In
  the closed ring geometry, the applied magnetic flux $f \ne 0$ adds additional
  relative rotation between the polarization axes at two ends of the same
  region.}

This topological difference leads to two important consequences which are
experimentally detectable.
First, the overlap between Majorana fermions has different nature according to
whether they are coupled through the NS or the TS regions. For example, in the
TS region, the pseudo-spins of two modes are always aligned and rotate in the
same way so they are always in phase, which is the main reason why the overlap
of Majorana states through the TS region is almost a constant independent of
phases. On the other hand, the two modes in the NS region are not aligned so
that their amplitudes depend on the superconducting phases and the magnetic
flux. In the following sections, we will see a stark contrast in the properties
of the supercurrents in the two cases.

Second, the two modes acquire different phases while they travel through the NS
region; see \figref{fig:topology}. While the $\nu=1$ mode in the left TS region
tunnels to the NS region without a rotation, it has to rotate by $2\gamma$ to
match with the $\nu=1$ mode in the right TS region. This rotation should be
reflected in the overlap matrix element between $\Psi_a$ and $\Psi_b$. However,
the $\nu=2$ mode rotates in the left TS-NS junction, while it is then already
aligned with the $\nu=2$ mode in the right TS region. The pseudo-spin rotation
at the left junction just contribute to an overall phase of $\Psi_a$ so that it
does not affect the $\Psi_a$-$\Psi_b$ overlap matrix element. Hence, the phase
difference $2\gamma$ between two modes arises.
Note that this additional phase does not take place in the case of the crossed
Andreev reflection (Cooper pair splitting) where two electrons in the middle NS
region go in the opposite directions. In this case the pseudo-spin rotates in
the opposite directions for opposite-moving electrons so that the phases are
canceled out.  Hence, this pseudo-spin rotation in the particle-hole space
severely affects the relative amplitudes of currents due to the single-electron
tunneling and the crossed Andreev reflection through the TS-NS-TS junctions.

\subsection{Subgap States and Supercurrent in a Closed Ring}
\label{sec:closedring}

For finite-size segments between two junctions, as in our system of ring
geometry (see \figref{fig:model}), the wave functions of two localized Majorana
modes have a finite overlap, which gives rise to finite energies $\pm E_A$ of
subgap eigenstates with $|E_A| \le E_{\rm gap} = \min(E_{\rm gap}^T,E_{\rm
  gap}^N)$ where $E_{\rm gap}^\ell$ is the gap in each region ($\ell=T,N$).
The overlap $E_A$ depends exponentially on the ratio of the segment length
$L_\ell$ ($\ell=N,T$) to the localization lengths $\lambda_{\ell\nu}$ of the
Majorana states.  The effective low-energy Hamiltonian can be then written as
\begin{align}
  \label{eq:HM}
  H_M = E_A (2d^\dag d - 1),
\end{align}
where $d = (\gamma_a + i\gamma_b)/\sqrt2$ is the fermionic operator from the
Majorana fermion operators $\gamma_{a,b}$. The subgap eigenstates are then
labeled as $\ket0$ and $\ket1 \equiv d^\dag\ket0$. The supercurrent
corresponding to the eigenstate is then calculated by taking the derivative of
the energy: \footnote{For the supercurrent, in general, one has to calculate
  all the states including the (almost) continuum of the bulk states. In our
  study we focus on the contribution from the subgap states.}
\begin{align}
  \label{eq:I}
  I = \frac{2e}{\hbar} \pde{H_M}{\delta\varphi}.
\end{align}

We determine the exact subgap energy $E_A$ by solving the BdG equation,
\eqnref{eq:BdGeq} in each region and matching the solutions across the
interfaces at $x=x_a$ and $x_b$ imposing the boundary conditions analogous to
\eqnref{eq:bc}. Explicitly, one has to solve self-consistently $E_A =
E_+(\tilde{k})$ [$E_A = E_-(\tilde{k})$ gives identical results due to the
particle-hole symmetry] and the boundary condition
\begin{align}
  \Psi(x_i^+) = \Psi(x_i^-) \,,\quad
  v_x\Psi(x_i^+) = v_x\Psi(x_i^-)
\end{align}
for $i=a,b$ and
\begin{align}
  \Psi(x)
  =
  \sum_{\eta\nu}
  \begin{cases}
    c_{\eta\nu}^N \chi^N_{k^N_{\eta\nu},+}(x) & (x_a < x < x_b)
    \\
    c_{\eta\nu}^T \chi^T_{k^T_{\eta\nu},+}(x) & (x_b < x < L)
  \end{cases}\ .
\end{align}
Here $k^\ell_{\eta\nu}$ are four solutions of $E_A = E_+(\tilde{k})$ with
$\tilde\mu = \tilde\mu_\ell$ and $c^\ell_{\eta\nu}$ the coefficient for each
mode. In the following section, the self-consistent equations are numerically
solved to obtain and examine the energy $E_A$ as a function of the magnetic
flux $f$ and the phase difference $\delta\varphi$ for given parameters.

Throughout the paper, we choose $R \approx 300\unit{nm}$, $m \approx 0.015 m_e$
($m_e$ is the bare electron mass), $\alpha \approx 2\times10^{-11} \unit{eVm}$,
and $\Delta_0\approx V_Z\approx 300\unit{\mu eV}$, which are suitable for
realistic samples. They correspond to $E_R \approx 20\unit{\mu eV}$ and
$\tilde\Delta\approx 3$ in the effective model.  The value of $\tilde\mu_\ell$
can be varied by the gate voltage.


\section{Results and Discussions}
\label{sec:results}

In this section we present the subgap eigenenergy $E_A$ and the supercurrent
$I$ by using the method described in \secref{sec:model}. First, we consider two
extreme cases where either the NS (\Secref{sec:shortNS}) or TS
(\Secref{sec:shortTS}) region is short compared with the localization lengths
of the Majorana states, and show that the two cases exhibit distinct behaviors
in subgap energy and supercurrent.
We then examine the evolution of one case to the other by changing continuously
the relative length between two regions, and the length dependence of each
supercurrent is discussed (\Secref{sec:length}).
Finally, we study the small ring case (\Secref{sec:smallRings}) where both the
behaviors should arise simultaneously.

\subsection{Short NS Region ($L_N\sim\lambda_{N1}$, $L_T\gg\lambda_{T1}$)}
\label{sec:shortNS}

\begin{figure}[t]
  \centering
  \includegraphics[width=7cm]{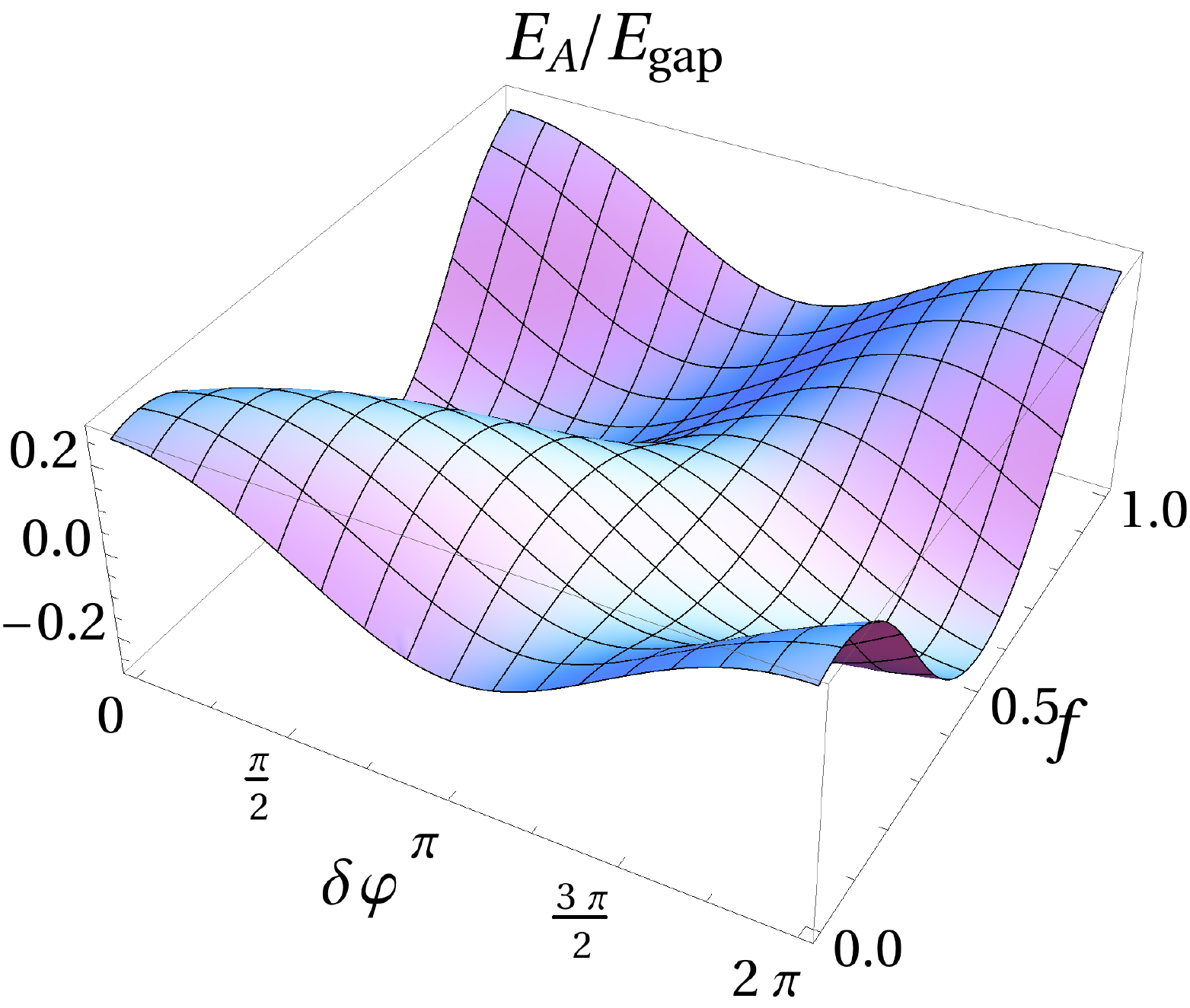}
  \caption{(Color online) Subgap energy for the state $\ket{1}$ as a function
    of $\delta\varphi$ and $f$ for the short NS region and long TS region:
    $L_N/\lambda_{N1} \approx 1.55$ and $L_T/\lambda_{T1} \approx 7.11$.  Here
    we have used $\tilde\Delta = 3$, $\tilde\mu_N = -5$, $\tilde\mu_T = 5$, and
    $E_\mathrm{gap} \approx 79\mu\unit{eV}$.}
  \label{fig:subgapE:LN200}
\end{figure}

\begin{figure}[t]
  \centering
  \includegraphics[width=8cm]{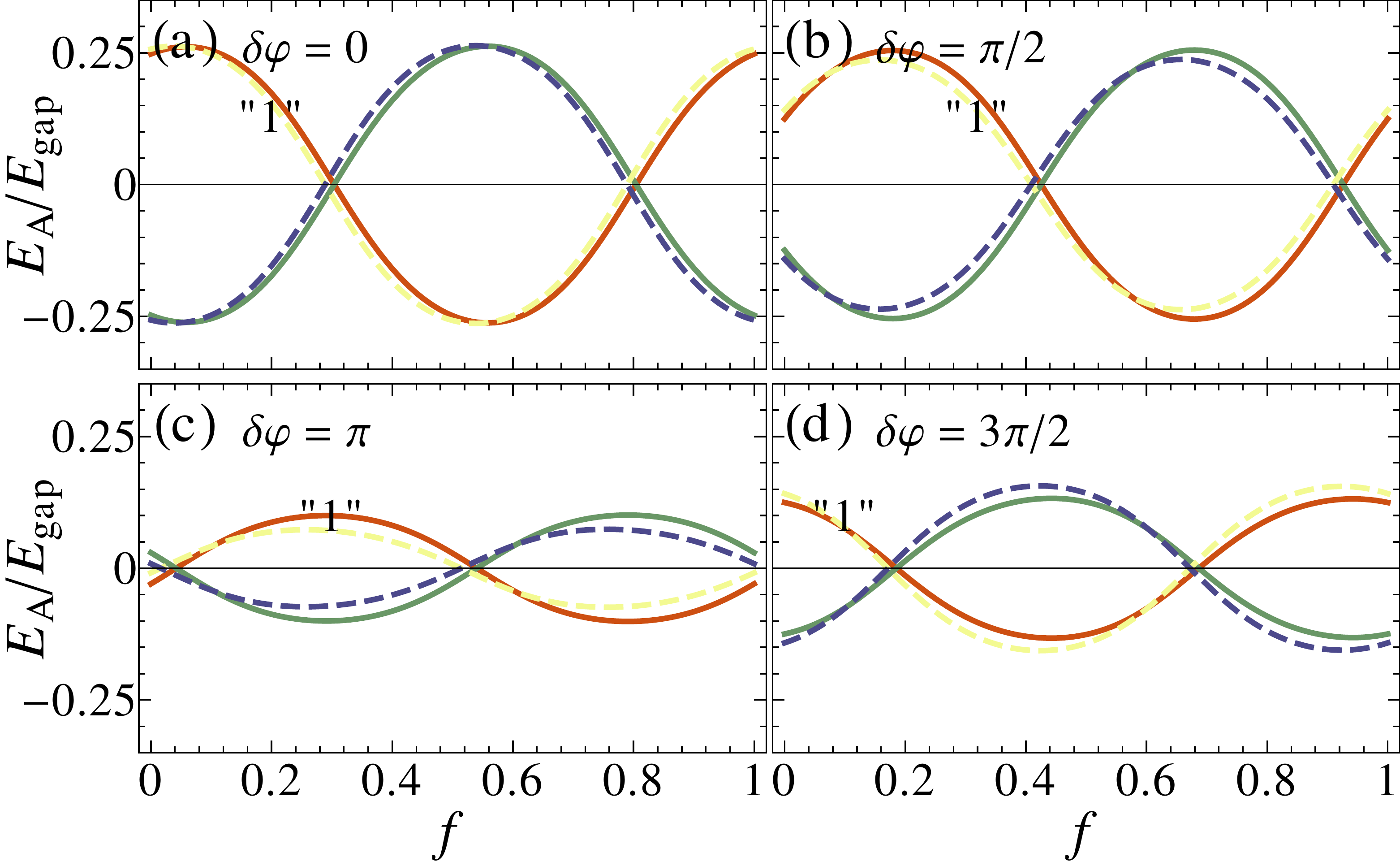}
  \caption{(Color online) Subgap energies as a function of $f$ for fixed values
    of $\delta\varphi$ as annotated. The energy which corresponds to the state
    $\ket{1}$ is marked by ``1''. Solid and dashed lines correspond to the
    exact and perturbative energies, respectively. We have used the same values
    for parameters as used in \figref{fig:subgapE:LN200}.}
  \label{fig:subgapE:LN200:dphi}
\end{figure}

\begin{figure}[t]
  \centering
  \includegraphics[width=8cm]{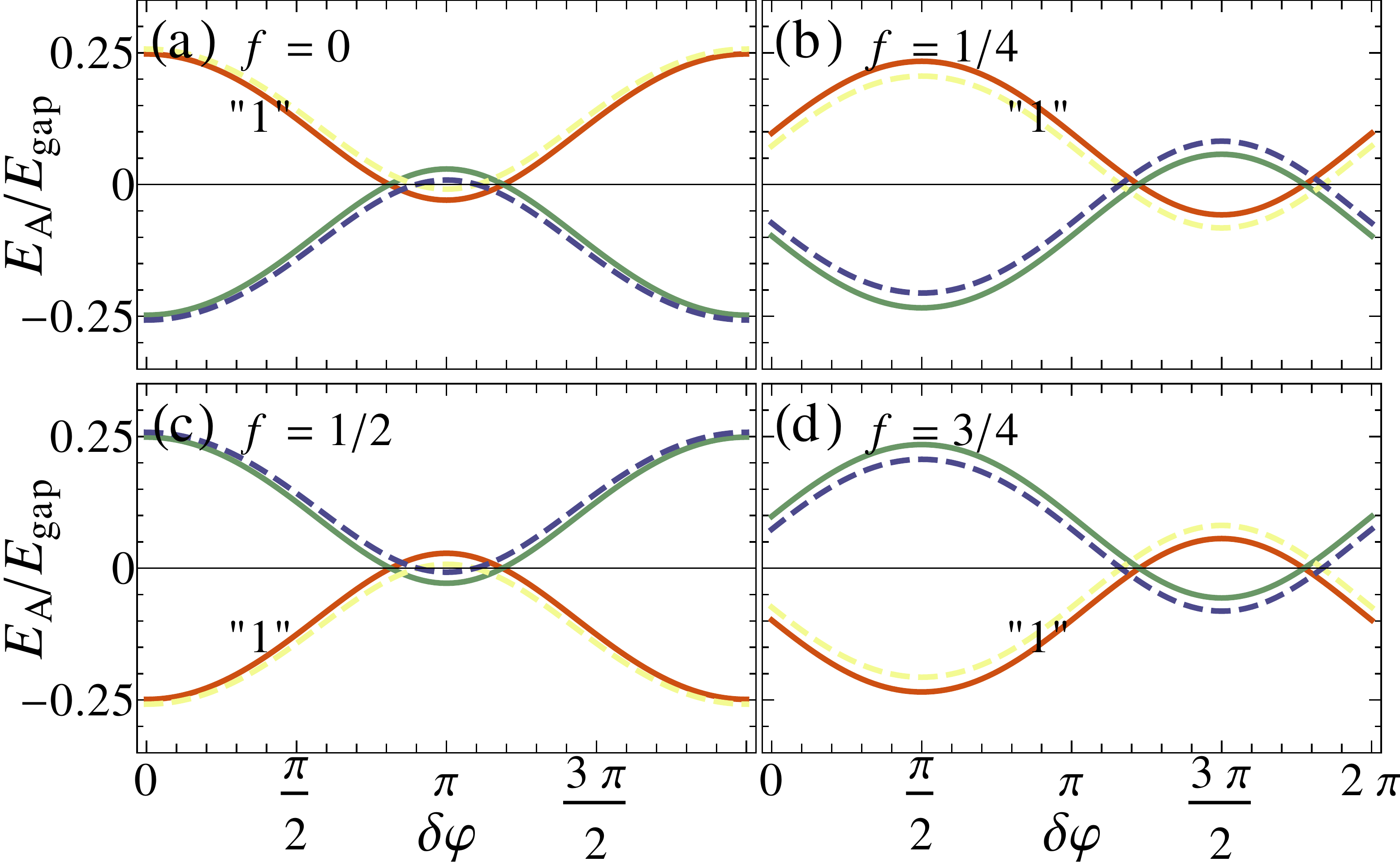}
  \caption{(Color online) Subgap energies as a function of $\delta\varphi$ for
    fixed values of $f$ as annotated. Same notations and parameters are used as
    in \figref{fig:subgapE:LN200:dphi}.}
  \label{fig:subgapE:LN200:f}
\end{figure}

First, we consider the case in which the NS region is short and the TS region
is very long: $L_N \sim \lambda_{N1}$ and $L_T \gg
\lambda_{T1}$. \Figsref{fig:subgapE:LN200}, \ref{fig:subgapE:LN200:dphi}, and
\ref{fig:subgapE:LN200:f} present our numerical results for the subgap
eigenenergy $E_A$ as functions of $\delta\varphi$ and $f$. We find that these
results fit well to the expression
\begin{align}
  \label{eq:EA:ns}
  E_A
  \approx
  E_M \cos (2\pi f + \gamma_M)
  + E_Z \cos(\delta\varphi - 2\pi f + \gamma_Z).
\end{align}
\Figsref{fig:subgapE:LN200:dphi} and \ref{fig:subgapE:LN200:f} show that for
$L_N \gtrsim \lambda_{N1}$, the coefficients $E_M$ and $E_Z$ and the phase
shifts $\gamma_M\approx\pi-2\gamma$ and $\gamma_Z\approx0$ are in a good
agreement with the approximate results (dashed line) calculated perturbatively
in \secref{sec:perturbation}.
In a large ring $(\tilde\Delta\gg1)$ with a short NS segment
($L_N\sim\lambda_{N1}$), the perturbation theory suggests a simpler expression
of the coefficients and the phase shifts: $\gamma_M\approx\gamma_Z\approx0$ and
\begin{subequations}
  \begin{align}
    \label{eq:EM:pt}
    E_M
    & \approx
    E_R \frac{\epsilon_1}{N_0} \left(e^{-L_N/\lambda_{N1}} + e^{-L_N/\lambda_{N2}}\right)
    \\
    \label{eq:EZ:pt}
    E_Z
    & \approx
    E_R \frac{\epsilon_1}{N_0} \left(e^{-L_N/\lambda_{N1}} - e^{-L_N/\lambda_{N2}}\right)
  \end{align}
\end{subequations}
Refer the definition of $\epsilon_1$ and $N_0$ to \eqnsref{eq:h} and
(\ref{eq:N}).

The subgap expression, \eqnref{eq:EA:ns} is consistent with that of Jiang et
al. \cite{Jiang11a} [see \eqnsref{eq:EMterm} and (\ref{eq:EZterm})]. To see
this, substitute the superconducting phases as follows:
\begin{align}
  \label{eq:substitution}
  \varphi_L & \to \varphi_T,
  &
  \varphi_M & \to \varphi_N,
  &
  \varphi_R & \to \varphi_T + 4\pi (f + 1/2).
\end{align}
Note that the phase shift $4\pi(f+1/2)$ in $\varphi_R$ is the phase
acquired by a Cooper pair circling around the ring in the presence of the
magnetic flux and the Rashba field. Then, the single-electron tunneling term
becomes
\begin{align}
  E_M^\mathrm{Jiang} \cos\frac{\varphi_L-\varphi_R}{2}
  \to
  - E_M^\mathrm{Jiang} \cos 2\pi f
\end{align}
corresponding to the $E_M$ term in \eqnref{eq:EA:ns}, and the Cooper pair
splitting term becomes
\begin{align}
  \label{eq:EZ:Jiang}
  E_Z^\mathrm{Jiang}
  \cos\left(\frac{\varphi_L+\varphi_R}{2} - \varphi_M\right)
  \to
  - E_Z^\mathrm{Jiang} \cos (\delta\varphi - 2\pi f).
\end{align}
corresponding to the $E_Z$ term in \eqnref{eq:EA:ns}.  The sign change is
ascribed to the Rashba phase which adds additional phase $\pi$ upon circling
around the ring.


The $E_M$-term comes from the circulation of a single electron around the
ring. In fact, the phase $2\pi f$ is exactly the magnetic phase acquired by a
single electron enclosing the magnetic flux $f$. The additional phase
$\gamma_M$ arises from the asymmetry between clockwise and counterclockwise
movers due to the finite curvature of the ring, as discussed in
\secref{sec:bulk}. \Eqnref{eq:EM:pt} shows that the contributions from two
modes are simply additive. As discussed in \secref{sec:topology}, the $\nu=1$
mode acquires the phase $2\gamma \approx \pi$ with respect to the $\nu=2$ mode,
which leads to a sign difference between them. On the other hand, the diagonal
component of the velocity operator in \eqnref{eq:v} suggests that the
supercurrent measures the pseudo-spin current. Since the pseudo-spins of the
two modes in the NS region are opposite to each other (when
$\gamma\approx\pi/2$), their contribution to the supercurrent is opposite in
sign. Hence, gathering two sign changes, there is no sign difference between
the two modes.

The $E_Z$-term is due to the Cooper pair tunneling between TS and NS region,
accompanying the splitting of the Cooper pair. The two electrons of a Cooper
pair tunnel between two regions through the two TS-NS boundaries,
respectively. In other words, the crossed Andreev reflection takes place
without any normal Andreev reflection accompanied. This perfect CAR is due to
the interesting characteristic of the TS-NS junction as discussed in
\secref{sec:introduction}: no Cooper pair can tunnel directly across a single
TS-NS junction.

Here three remarks are worthwhile concerning the CAR process involved in the
$E_Z$-term.  (i) As seen in \eqnsref{eq:EA:ns} and
(\ref{eq:EZ:Jiang}), the CAR process acquires the phase $\delta\varphi - 2\pi
f$. The phase $\delta\varphi$ is obviously due to the tunneling of a Cooper
pair between two different superconductors. The appearance of the phase $2\pi
f$ is interesting because it is identical to that by the circulation of a
single electron. It indicates that the splitting and the recombination of the
Cooper pair should take place at the same TS-NS boundary so that only one of
two electrons split moves around the ring before the recombination, resulting
in the phase $2\pi f$. The recombined Cooper pair at one of the boundaries
then flows into the bulk superconductor, not being affected by the magnetic
flux any more. It is consistent with the fact that the Majorana fermions are
localized at the boundaries. This dependence on the magnetic flux is the
evidence that the CAR is realized via the Majorana fermions.

\begin{figure}[t]
  \centering
  \includegraphics[width=7cm]{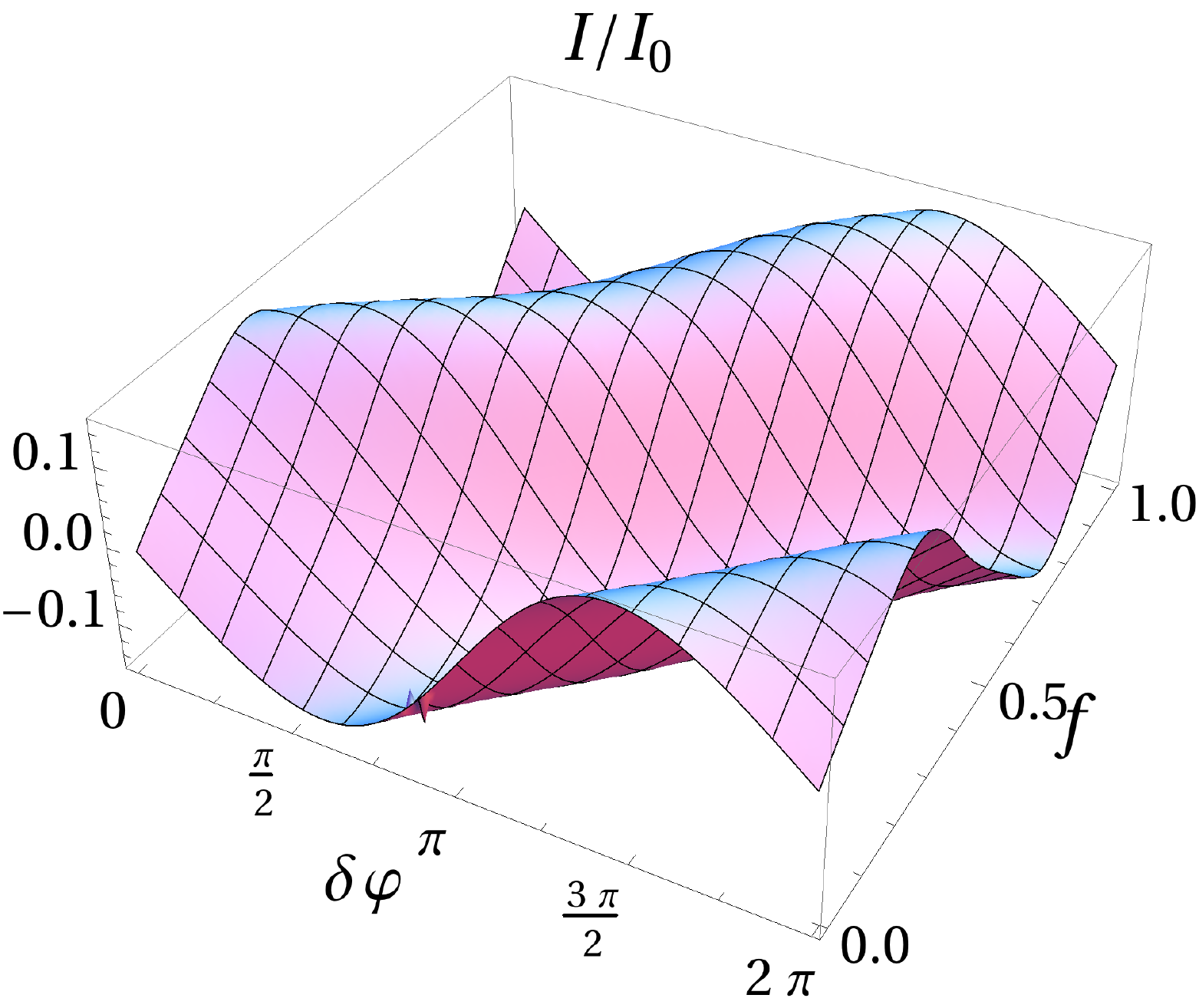}
  \caption{(Color online) Supercurrent for the state $\ket{1}$ as a function of
    $\delta\varphi$ and $f$ for the short non-topological region and long
    topological region. We have used the same values for parameters as used in
    \figref{fig:subgapE:LN200}. Here $I_0 \equiv \frac{2e}{\hbar}
    E_\mathrm{gap}$.}
  \label{fig:I:LN200}
\end{figure}

(ii) Unlike in the $E_M$-term above, no extra pseudo-spin rotation between two
modes (see \Secref{sec:topology}) accompanying tunneling across NS-TS junctions
takes place, while the pseudo-spin currents of the two modes are opposite in
sign. Hence, \eqnref{eq:EZ:pt} exhibits the negative combination between two modes.

(iii) As a consequence of the effect (ii), the $E_Z$-term depends
non-monotonically on the length $L_N$ of the NS segment.
The exponential factor $e^{-L_N/\lambda_{N\nu}}$ in both the $E_M$- and
$E_Z$-term implies that the finite overlap between Majorana fermions is
indispensable to observe these processes. It is also known that the CAR
process can happen substantially only over lengths shorter than the size of
the Cooper pair (i.e., the superconducting coherence length).
Based on both, one may naively expect that the CAR process (and hence $E_Z$)
get stronger with decreasing $L_N$. However, the tunneling processes through
two modes $\nu=1,2$ gives opposite contributions as shown in the above (ii),
due to the topological characteristic of the subgap states. Therefore the CAR
process becomes weaker if the NS segment is too small: $E_Z$ increases as $L_N$
decreases until $L_N\geq\lambda_{N2}$, but decreases again if $L_N$ decreases
further and gets smaller than $\lambda_{N2}$ ($\leq\lambda_{N1}$).
This non-monotonic dependence of the $E_Z$-term on $L_N$ will indeed be
demonstrated explicitly in \Figsref{fig:L:st}(a) and \ref{fig:L:vst_wt}(a),
\secref{sec:length}.

Since in both the $E_M$- and $E_Z$-term it is a single electron, not a Cooper
pair, that circulates around the ring, the periodicity of the subgap energy
$E_A$ with respect to the magnetic flux $f$ is 1, not 1/2 as in the normal
superconductor ring, which is clearly revealed in
\figref{fig:subgapE:LN200:dphi}. The $f = 1$ periodicity is protected as long
as no fermion-parity breaking mechanism is introduced into the system; If the
parity breaking is present, the periodicity would be reduced to 1/2.

\begin{figure}[t]
  \centering
  \includegraphics[width=8cm]{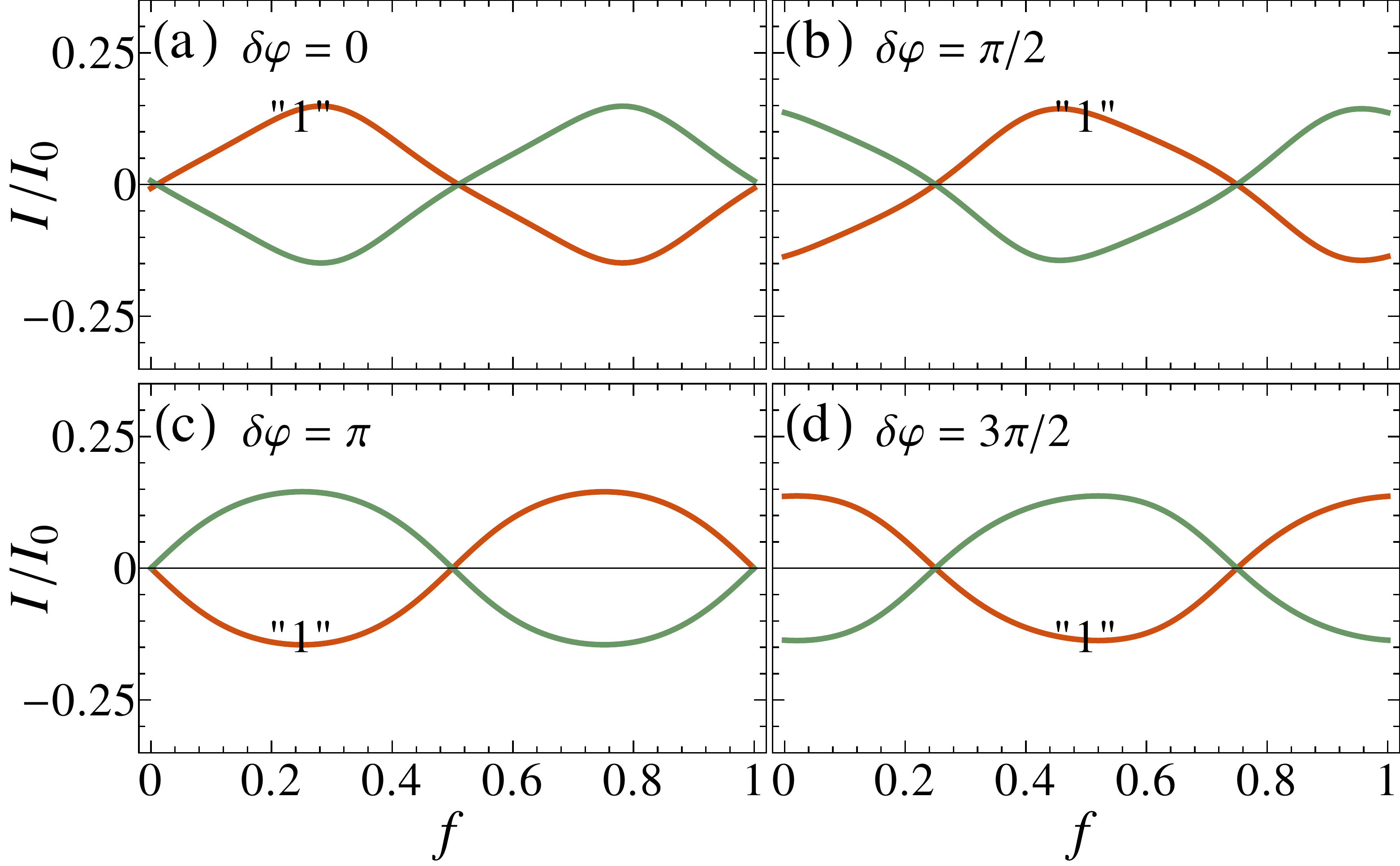}
  \caption{(Color online) Supercurrent as a function of $f$ for fixed values of
    $\delta\varphi$ as annotated. Same notations and parameters are used as
    in \figref{fig:subgapE:LN200:dphi}.}
  \label{fig:I:LN200:dphi}
\end{figure}

\begin{figure}[t]
  \centering
  \includegraphics[width=8cm]{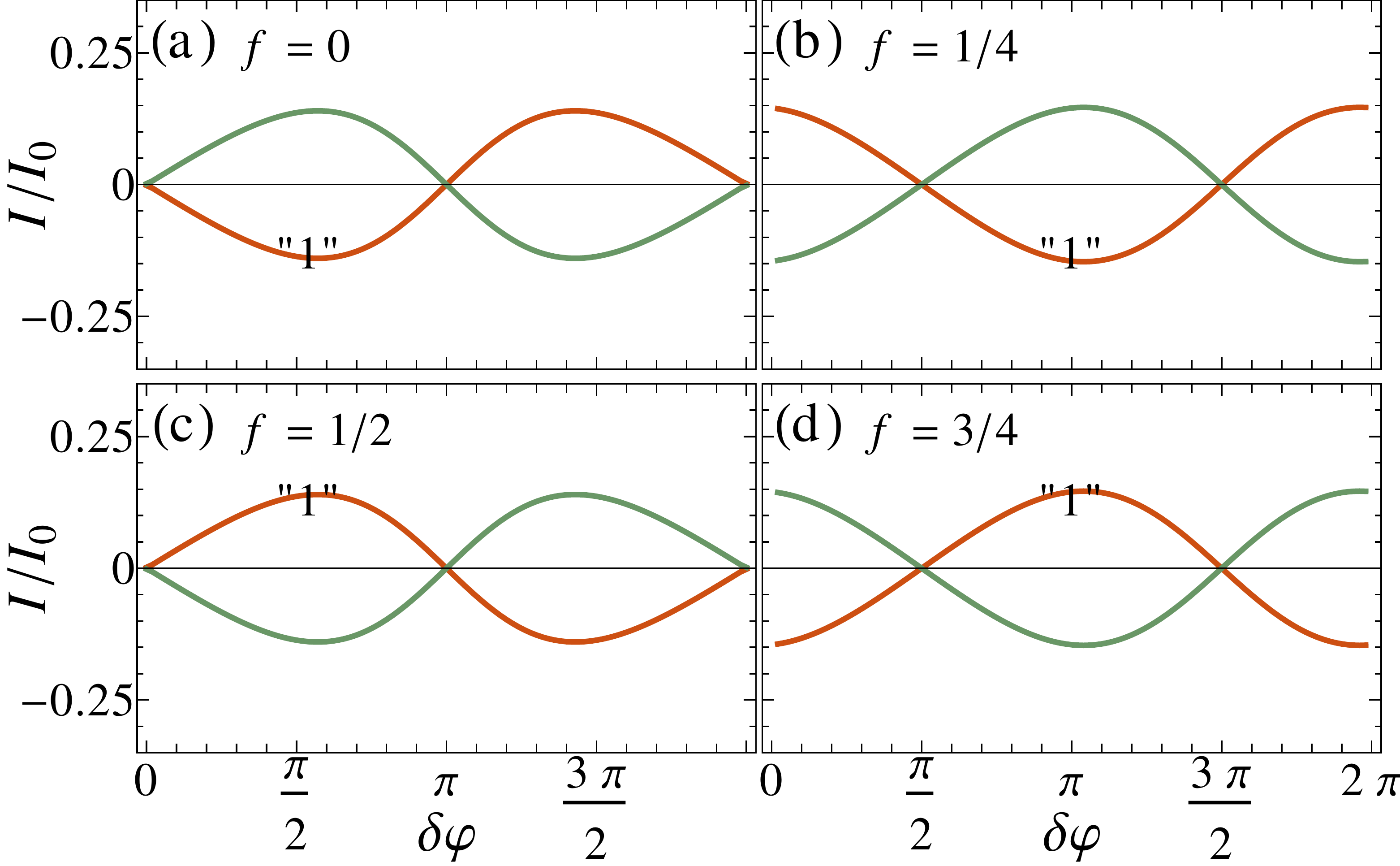}
  \caption{(Color online) Supercurrent as a function of $\delta\varphi$ for
    fixed values of $f$ as annotated. Same notations and parameters are used as
    in \figref{fig:subgapE:LN200:dphi}.}
  \label{fig:I:LN200:f}
\end{figure}

\Figsref{fig:I:LN200}, \ref{fig:I:LN200:dphi}, and \ref{fig:I:LN200:f} show the
corresponding supercurrent, assuming that the fermion parity is preserved. From
\eqnsref{eq:HM}, (\ref{eq:I}), and (\ref{eq:EA:ns}), the supercurrent is
obtained as
\begin{align}
  \label{eq:I:ns}
  I
  \approx (2d^\dag d - 1) \frac{2e}{\hbar} E_Z \sin(\delta\varphi - 2\pi f).
\end{align}
The supercurrent obtained numerically is not exactly sinusoidal since it
includes the contribution from higher-order processes: In fact, the
coefficients $E_Z$ and $E_M$ are also functions of $\delta\varphi$ through the
normalization constants $N_a$ and $N_b$ [see \secref{sec:perturbation}].  As a
matter of fact, only the $E_Z$-term contributes to the supercurrent: the
$E_M$-term does not involve the transport of the Cooper pair. The current also
exhibits the $f=1$ periodicity, which is the fingerprint of the Majorana
fermions.
In the presence of the parity breaking, the $f=1$ periodicity may fade
away. However, one can still detect the existence of the Majorana fermion by
examining the response of the supercurrent with respect to the variation of
both $\delta\varphi$ and $f$. In the following section, we will compare the
supercurrents due to the crossed and normal Andreev reflections and discuss how
to distinguish them. One thing to be noted here is that in the CAR process the
role of the magnetic flux $f$ is shifting the current by $2\pi f$ without
modulating the amplitude of the current.

Finally, we would like to note that all the properties of the subgap energy and
the supercurrent are independent of the length of the TS region as long as it
is sufficiently larger than the size of the Majorana fermions. It is in
contrast to the high dependence of the $E_Z$-term on the relative length
between the NS segment length and the Cooper pair size. It indicates that the
Majorana fermion state in the TS region is highly nonlocal. In other words,
this $L_T$-independence reflects that the correlation length and the size of
the Cooper pair in the TS are almost infinite as long as the coherence is
preserved.

\subsection{Short TS Region ($L_N\gg\lambda_{N1}$, $L_T\sim\lambda_{T1}$)}
\label{sec:shortTS}

\begin{figure}[b]
  \centering
  \includegraphics[width=7cm]{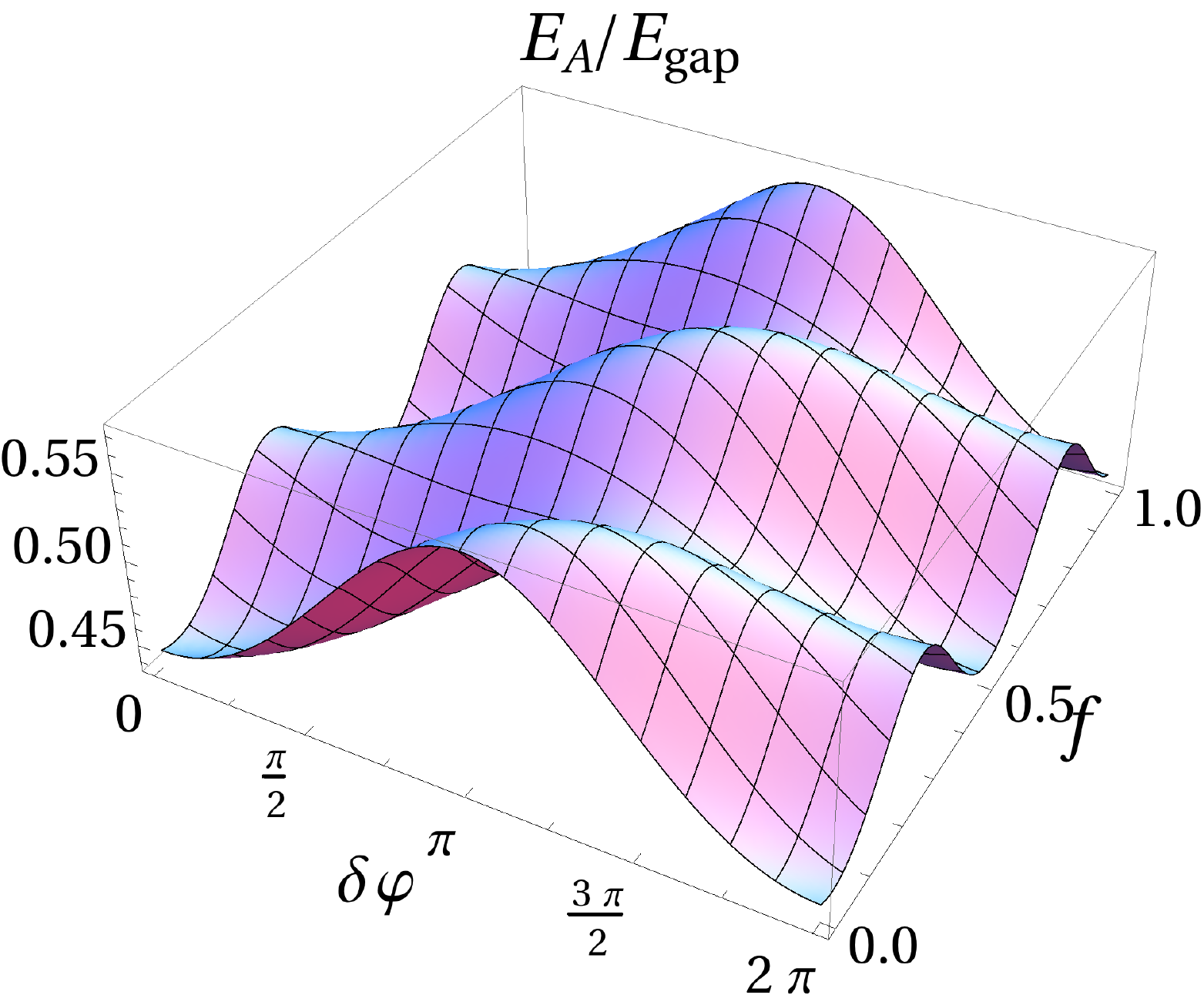}
  \caption{(Color online) Subgap energy for the state $\ket{1}$ as a function
    of $\delta\varphi$ and $f$ for the long non-topological region and short
    topological region: $L_N/\lambda_{N1} \approx 6.96$ and $L_T/\lambda_{T1}
    \approx 0.89$.  Here we have used $\tilde\Delta = 3$, $\tilde\mu_N = -5$,
    $\tilde\mu_T = 5$, and $E_\mathrm{gap} \approx 79\mu\unit{eV}$.}
  \label{fig:subgapE:LN900}
\end{figure}

\begin{figure}[t]
  \centering
  \includegraphics[width=8cm]{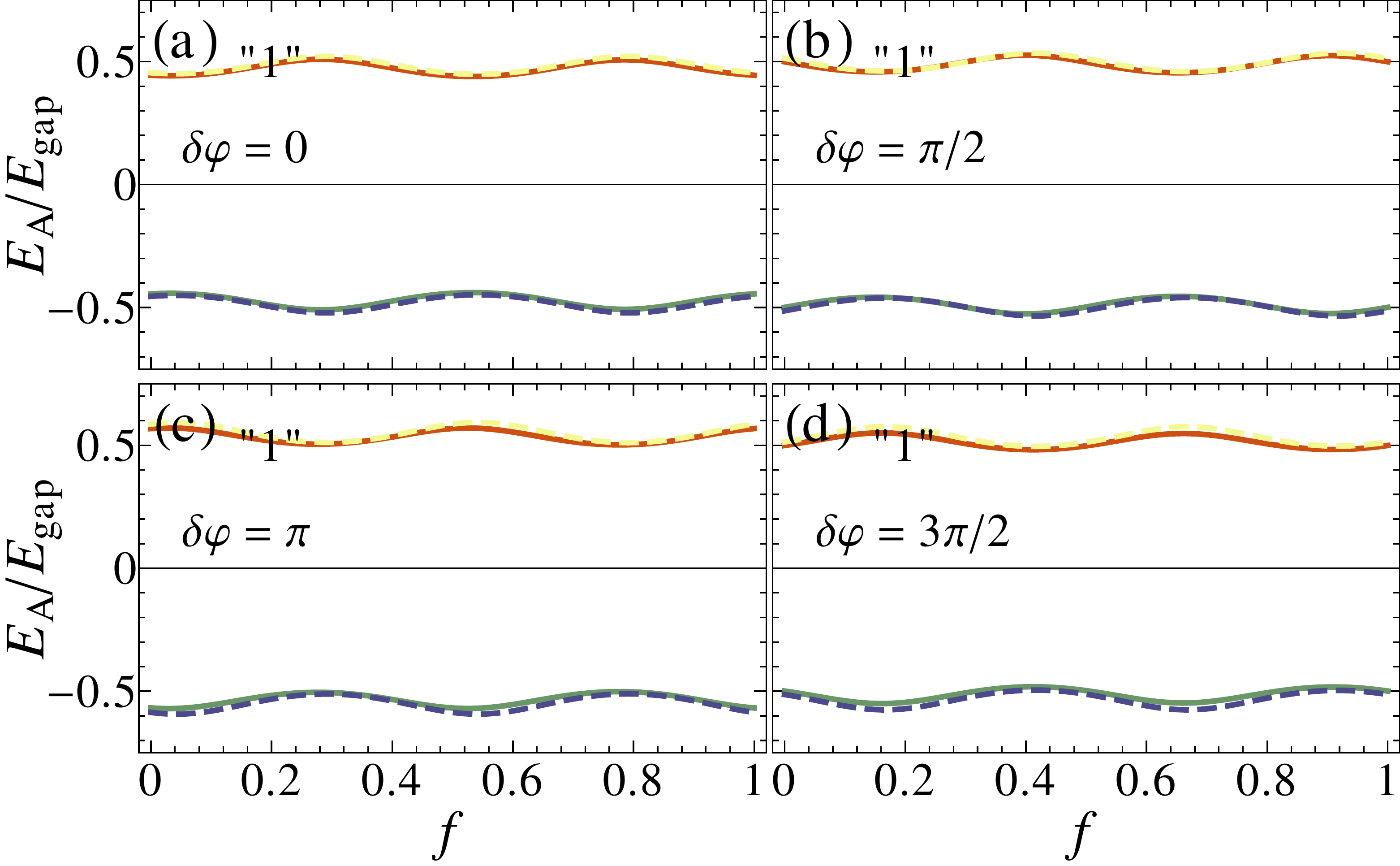}
  \caption{(Color online) Subgap energies as a function of $f$ for fixed values
    of $\delta\varphi$ as annotated. We have used the same values for
    parameters as used in \figref{fig:subgapE:LN900}.}
  \label{fig:subgapE:LN900:dphi}
\end{figure}

\begin{figure}[t]
  \centering
  \includegraphics[width=8cm]{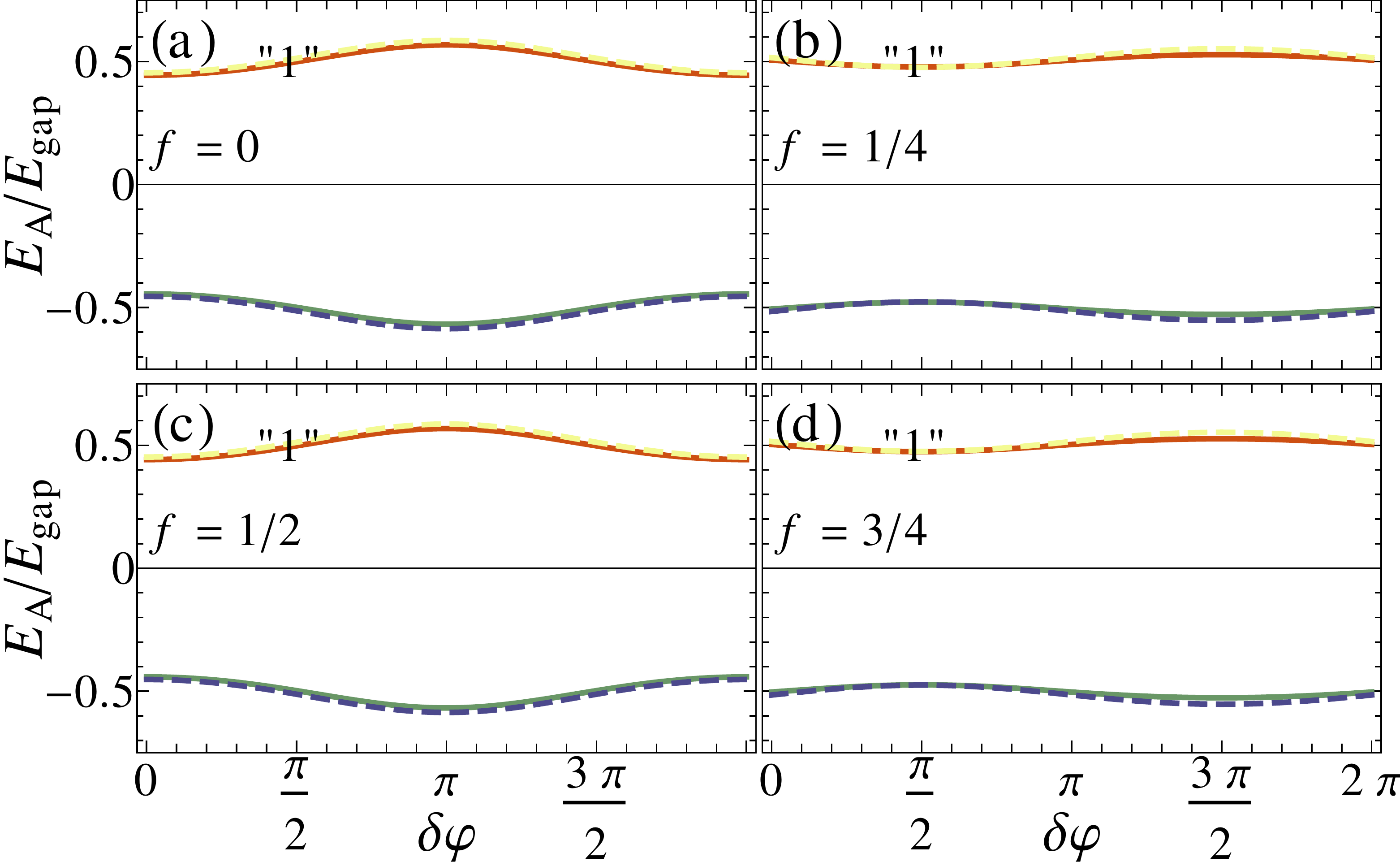}
  \caption{(Color online) Subgap energies as a function of $\delta\varphi$ for
    fixed values of $f$ as annotated. Same notations and parameters are used as
    in \figref{fig:subgapE:LN900:dphi}.}
  \label{fig:subgapE:LN900:f}
\end{figure}

Now we consider the opposite case in which the TS region is short and the NS
region long: $L_T\sim\lambda_{T1}$ and $L_N\gg\lambda_{N1}$. Interestingly, in
this case the physics of the Majorana fermions is completely different as shown
in \Figsref{fig:subgapE:LN900}, \ref{fig:subgapE:LN900:dphi}, and
\ref{fig:subgapE:LN900:f}. In this regime, we obtain the following empirical
expression for the subgap energy:
\begin{align}
  \label{eq:EA:ts}
  E_A
  \approx
  E_0 + E_C \left[\cos \delta\varphi + \cos(4\pi f - \delta\varphi)\right].
\end{align}
\Figsref{fig:subgapE:LN900:dphi}, and \ref{fig:subgapE:LN900:f} show that our
exact and perturbative results match well with each other. The simpler
expressions for the coefficients $E_0$ and $E_C$ are at hand in the large-ring
limit and for $L_T/\lambda_{T1} \gtrsim 1$:
\begin{subequations}
  \begin{align}
    \label{eq:E0:pt}
    E_0
    & \approx
    \frac{E_R}{N_0}
    \begin{cases}
      \displaystyle
      (\epsilon_1+\epsilon_2) e^{-L_T/\lambda_{T1}}
      &
      \\
      \displaystyle
      \quad\mbox{}
      + (\epsilon_1-\epsilon_2) e^{-L_T/\lambda_{T2}},
      & \tilde\mu_T < D
      \\
      \displaystyle
      2\epsilon_1 e^{-L_T/\lambda_{T1}} \cos k_r L_T
      &
      \\
      \displaystyle
      \quad\mbox{}
      + 2\epsilon_2 e^{-L_T/\lambda_{T1}} \sin k_r L_T,
      & \tilde\mu_T > D
    \end{cases}
    \\
    \label{eq:EC:pt}
    E_C
    & \approx \frac{\sqrt{D}}{2\tilde\mu_N N_0} E_0.
  \end{align}
\end{subequations}
See \eqnsref{eq:h} and (\ref{eq:N}) for the definition of $\epsilon_{1,2}$ and
$N_0$.

The overlap between the Majorana fermions through the TS region gives rise to a
finite constant level splitting, $E_0$-term which is independent of $f$ and
$\delta\varphi$. Note that such a constant term is missing in the former case
where the overlap happens in the NS region. This is attributed to the
topological difference between subgap states in TS and NS regions: The
pseudo-spin directions of the two subgap states ($\nu=1,2$) in the TS region
are parallel to each other (see \Figref{fig:topology}). Technically, the
coefficients in \eqnsref{eq:ca} and (\ref{eq:cb}) in the TS region do not
depend on the phases $f$ and $\delta\varphi$. The constant splitting $E_0$
increases as the TS segment length decreases, eventually reaching the band gap
$E_\mathrm{gap}$ at $L_T\to0$. Because of this constant splitting, no crossing
between the subgap states at the Fermi level takes place.

The phase-dependent term, $E_C$-term is identical to that of a SQUID
made of two normal Josephson junctions threaded by a magnetic flux $f$, in
which the phase differences in the two junctions are $\delta\varphi$ and $4\pi
f - \delta\varphi$, respectively. The $E_C$-term can be directly inferred by
substituting the superconducting phases in \eqnref{eq:ECterm} according to the
same rule, \eqnref{eq:substitution} as used in the short-NS-region case.
\begin{align}
  \begin{split}
    & E_C
    \left[
      \cos(\varphi_L-\varphi_\mathrm{M})
      + \cos(\varphi_\mathrm{M}-\varphi_R)
    \right]
    \\
    & \ \to\
    E_C \left[\cos\delta\varphi + \cos(4\pi f - \delta\varphi)\right]
  \end{split}
\end{align}
This dependence on $\delta\varphi$ and $f$ confirms our prediction discussed in
\secref{sec:introduction} that it is a Cooper pair that tunnels through the
NS-TS junctions if the Majorana fermions are coupled via the NS region [see
\figref{fig:doublejunctions}(b)]. The overlap between Majorana fermions opens a
channel at $E_A\ne0$. The Cooper pair then circulates around the ring via the
successive Andreev reflection in each junction.  Hence the periodicity of the
subgap energy $E_A$ with respect to the magnetic flux $f$ is 1/2 as can be seen
in \figref{fig:subgapE:LN900:dphi}. Note that the Rashba phase does not affect
the $E_C$-term since it gives rise to a phase $4\pi\times1/2 = 2\pi$. Namely, the Rashba phase acquired by a Cooper pair is twice larger than that of
a single electron.

\begin{figure}[b]
  \centering
  \includegraphics[width=7cm]{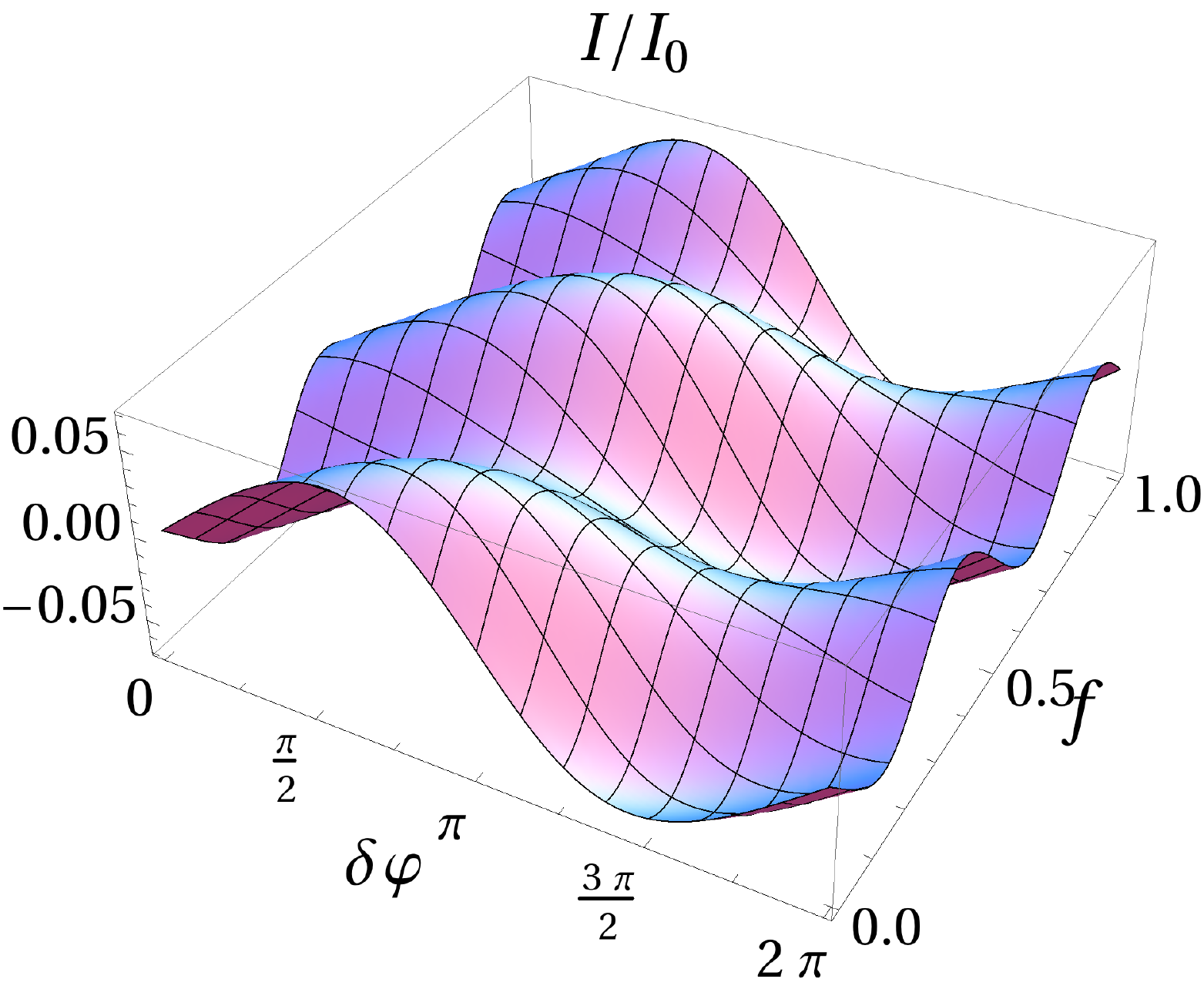}
  \caption{(Color online) Supercurrent for the state $\ket{1}$ as a function of
    $\delta\varphi$ and $f$ for the long non-topological region and short
    topological region. We have used the same values for parameters as used in
    \figref{fig:subgapE:LN900}. Here $I_0 \equiv \frac{2e}{\hbar}
    E_\mathrm{gap}$.}
  \label{fig:I:LN900}
\end{figure}

\begin{figure}[t]
  \centering
  \includegraphics[width=8cm]{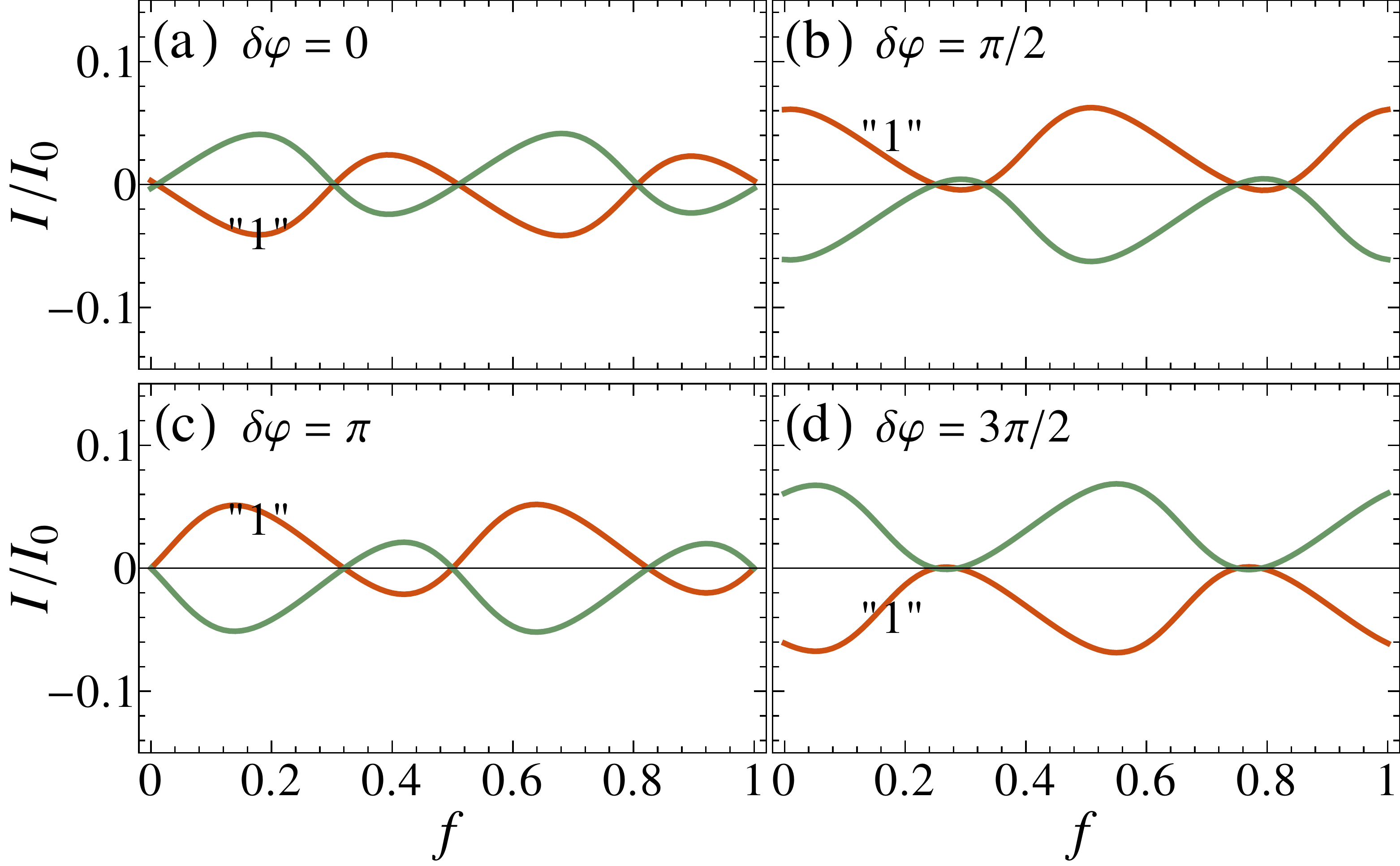}
  \caption{(Color online) Supercurrent as a function of $f$ for fixed values of
    $\delta\varphi$ as annotated. Same notations and parameters are used as
    in \figref{fig:subgapE:LN900:dphi}.}
  \label{fig:I:LN900:dphi}
\end{figure}

\begin{figure}[t]
  \centering
  \includegraphics[width=8cm]{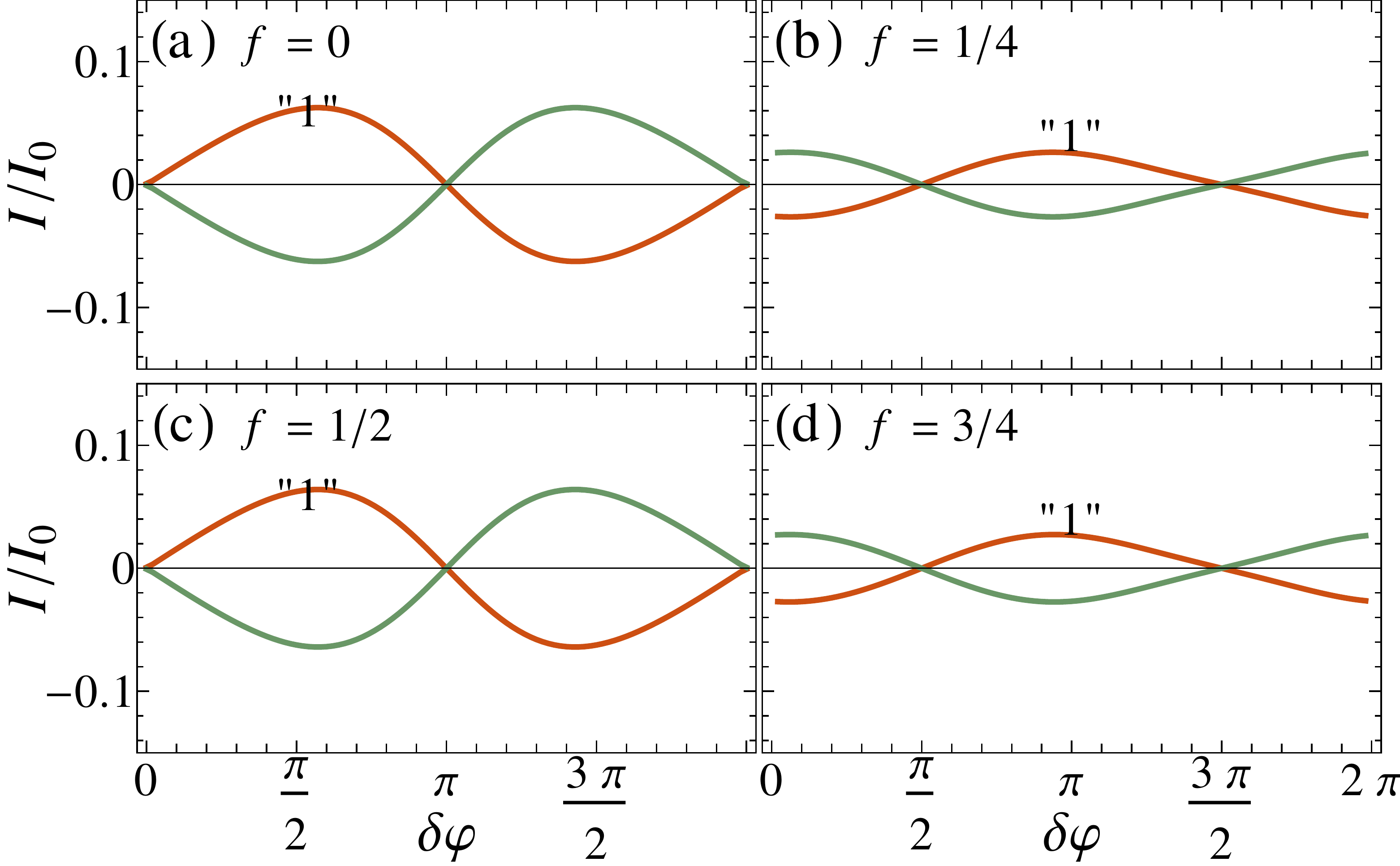}
  \caption{(Color online) Supercurrent as a function of $\delta\varphi$ for
    fixed values of $f$ as annotated. Same notations and parameters are used as
    in \figref{fig:subgapE:LN900:dphi}.}
  \label{fig:I:LN900:f}
\end{figure}

\Figsref{fig:I:LN900}, \ref{fig:I:LN900:dphi}, and \ref{fig:I:LN900:f} show the
corresponding supercurrent. From \eqnsref{eq:HM}, (\ref{eq:I}), and
(\ref{eq:EA:ts}), the supercurrent is obtained as
\begin{align}
  \label{eq:I:ts}
  \begin{split}
    I
    & \approx (1 - 2d^\dag d) \frac{2e}{\hbar} E_C
    \left[
      \sin\delta\varphi + \sin(\delta\varphi - 4\pi f)
    \right]
    \\
    & =
    (1 - 2d^\dag d) \frac{4e}{\hbar} E_C
    \cos2\pi f\sin(\delta\varphi - 2\pi f).
  \end{split}
\end{align}
The approximate expression, \eqnref{eq:I:ts} shows some discrepancy from the
numerically exact results. It is because the latter includes the contributions
from higher-order processes: The current does not vanish at $f=1/4$ and
$3/4$. The $f{=}1/2$ periodicity is well shown in \figref{fig:I:LN900:dphi}.
By comparing \eqnref{eq:I:ns} and (\ref{eq:I:ts}), one can notice that while in
both cases the magnetic flux $f$ shifts the current by $2\pi f$, it also
modulates clearly the amplitude of the current in the NAR process with the
weighting factor $\cos2\pi f$: see the variation of the current amplitudes with
respect to $f$ in \figref{fig:I:LN900:f}. Hence, apart from the periodicity
with respect to $f$, the modulation of the current can be used to detect the
CAR phenomena due to the Majorana fermions.

\Figsref{fig:I:LN900}, \ref{fig:I:LN900:dphi}, and \ref{fig:I:LN900:f} show the
corresponding supercurrent. From \eqnsref{eq:HM}, (\ref{eq:I}), and
(\ref{eq:EA:ts}), the supercurrent is obtained as
\begin{align}
  \label{eq:I:ts}
  \begin{split}
    I
    & \approx (1 - 2d^\dag d) \frac{2e}{\hbar} E_C
    \left[
      \sin\delta\varphi + \sin(\delta\varphi - 4\pi f)
    \right]
    \\
    & =
    (1 - 2d^\dag d) \frac{4e}{\hbar} E_C
    \cos2\pi f\sin(\delta\varphi - 2\pi f).
  \end{split}
\end{align}
The numerically calculated supercurrent does not exactly behave as
\eqnref{eq:I:ts} since it includes the contributions from higher-order
processes: The current does not vanish at $f=1/4$ and $3/4$. The
$f{=}1/2$ periodicity is well shown in \figref{fig:I:LN900:dphi}.
By comparing \eqnref{eq:I:ns} and (\ref{eq:I:ts}), one can notice that while in
both cases the magnetic flux $f$ shifts the current by $2\pi f$, it also
modulates clearly the amplitude of the current in the NAR process with the
weighting factor $\cos2\pi f$: see the variation of the current amplitudes with
respect to $f$ in \figref{fig:I:LN900:f}. Hence, apart from the periodicity
with respect to $f$, the modulation of the current can be used to detect the
CAR phenomena due to the Majorana fermions.

One more interesting property of the $E_0$- and $E_C$-term, in comparison with
the $E_Z$- and $E_M$-term, is that their sign can be controlled by tuning the
TS segment length and/or the chemical potential $\tilde\mu_T$ in the deep
topological region $(\tilde\mu_T>D)$ [see \eqnref{eq:EC:pt}]. Apart from the
exponentially decreasing envelop part, the $E_0$ and $E_C$-terms are
oscillatory with $k_r L_T$. This is owing to the oscillatory behavior of the
Majorana wave function in the deep topological phase due to the finite real
part of the wave vectors, $k_r$ [see \eqnref{eq:kr}]. This oscillatory feature
is peculiar in that it cannot be observed in the usual normal superconductor
SQUID hosting no Majorana fermions. It provides an electronic way to change the
sign of the supercurrent.

\subsection{Length Dependence}
\label{sec:length}

\begin{figure}[t]
  \centering
  \includegraphics[width=8cm]{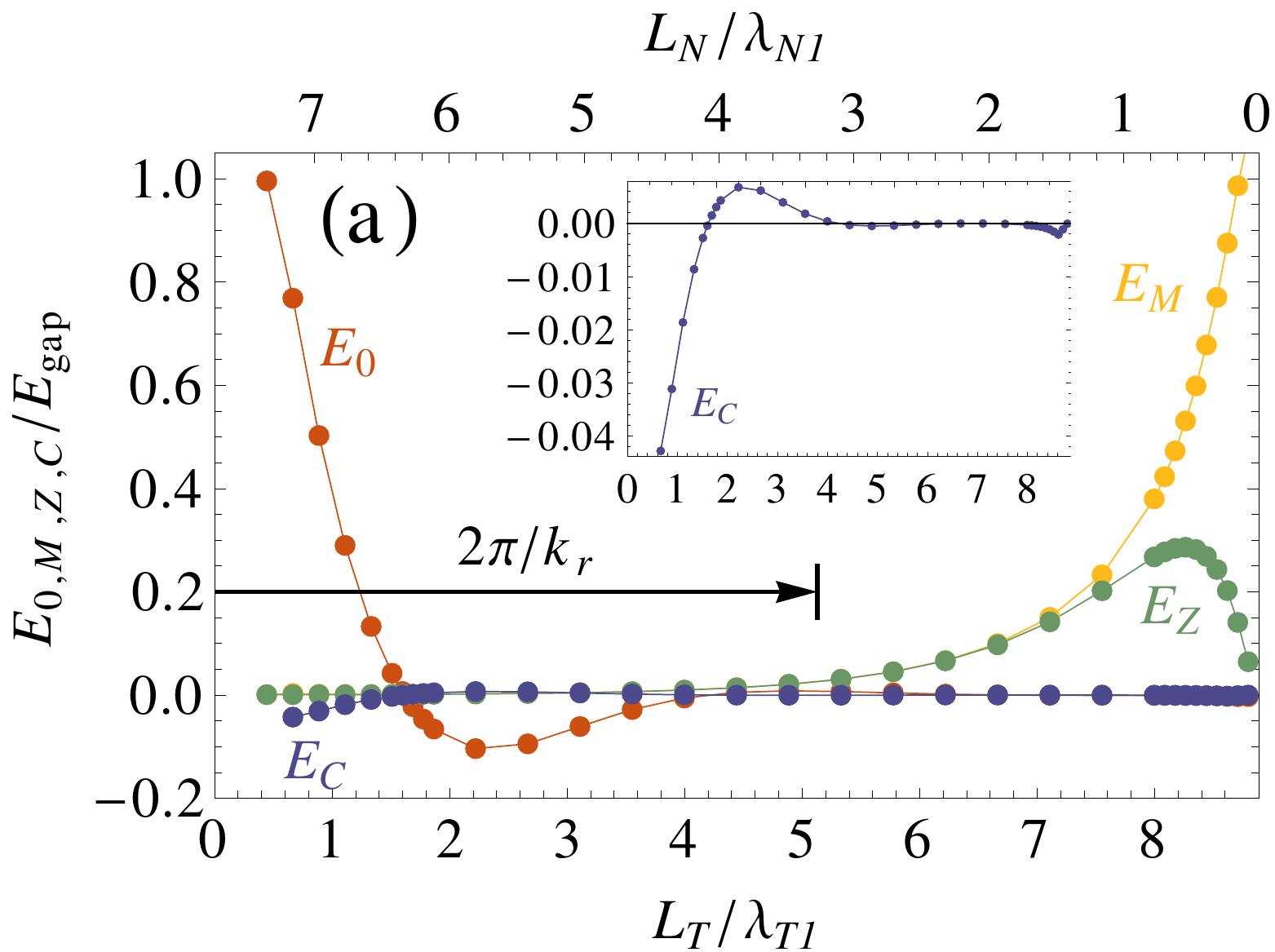}\\
  \includegraphics[width=8cm]{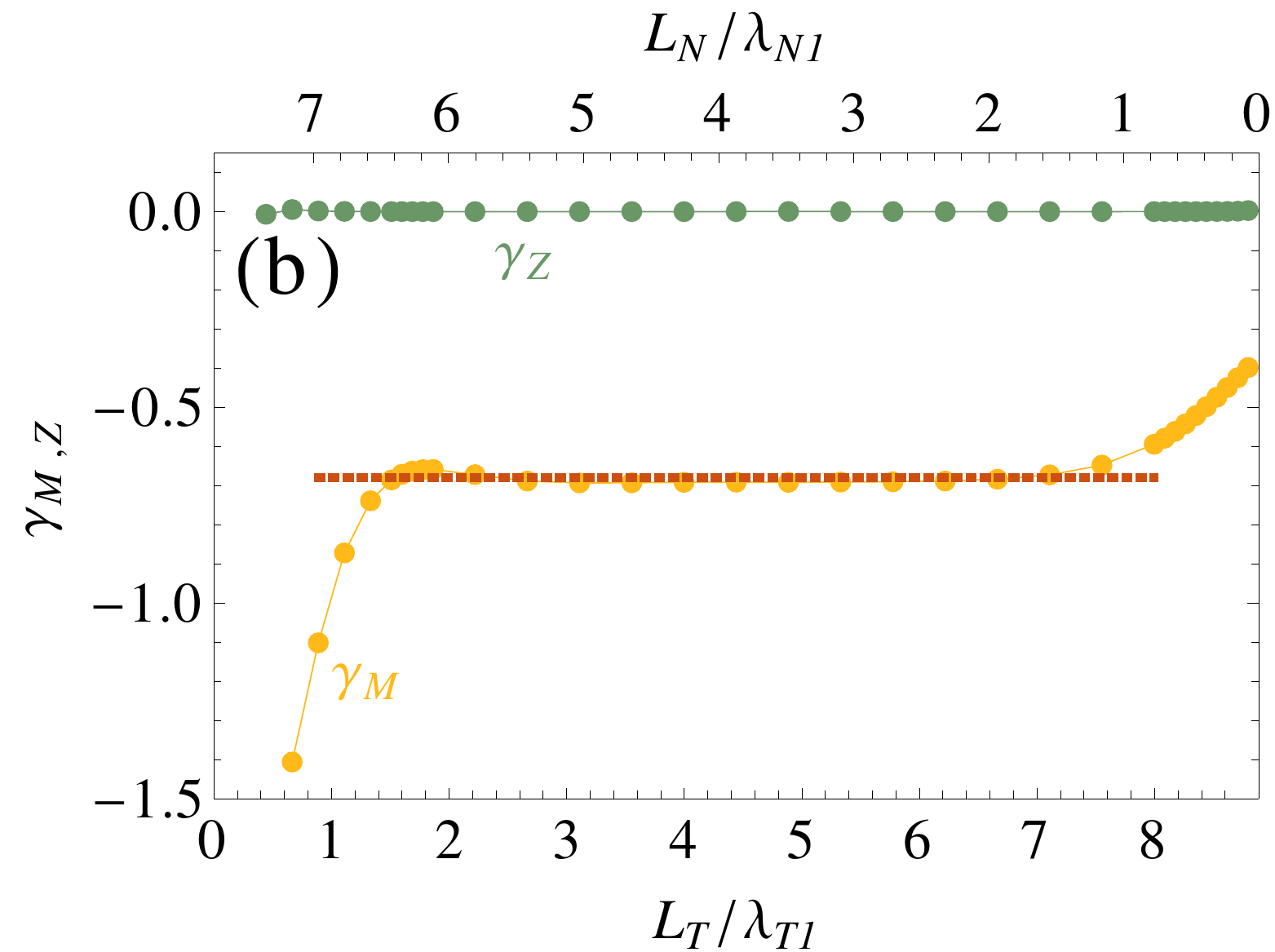}
  \caption{(Color online) (a) The coefficients $E_0$, $E_C$, $E_M$, and $E_Z$
    as functions of $L_N$ or $L_T=L-L_N$ with $L = L_N+L_T$ fixed. The inset
    shows the enlarged view of $E_C$. (b) The phase shifts $\gamma_M$ and
    $\gamma_Z$ as functions of $L_N$ or $L_T$. The dashed line correspond to
    the perturbative prediction for $\gamma_M$, $\pi - 2\gamma$. Here we have
    used $\tilde\Delta = 3$, $\tilde\mu_N = -5$, and $\tilde\mu_T = 5$.}
  \label{fig:L:st}
\end{figure}

Having understood the transport mechanisms in two extreme cases, we now
examine intermediate cases varying the relative lengths of the NS and TS
segment. \Figref{fig:L:st} displays the dependence of the coefficients ($E_0$,
$E_C$, $E_M$, and $E_Z$) and the phase shifts ($\gamma_M$ and $\gamma_Z$) on
the segment lengths $L_N$ and $L_T$ with $L=L_N+L_T$ fixed.
Obviously, the coefficients $E_M$ and $E_Z$, exhibiting the exponential
dependence on $L_N$, are finite for $L_N\lesssim\lambda_{N1}$, and $E_0$ and
$E_C$ are so for $L_T\lesssim\lambda_{T1}$. For $L_N\gg\lambda_{N1}$ and
$L_T\gg\lambda_{T1}$, all the coefficients are vanishingly small, since the
overlap between the Majorana fermions is negligible and no transport through
the NS-TS junction is possible.
The coefficients $E_M$ and $E_0$ approach the energy gap $E_\mathrm{gap}$ as
$L_N\to0$ and $L_T\to0$, respectively. In these limits, the Majorana fermions
are strongly bound so that they become completely fermionic.
On the other hand, the coefficients $E_Z$ and $E_C$, responsible for the
tunneling of Cooper pairs through the ring, remain relatively small as the
segments length decreases. $E_Z$ goes to zero as $L_N\to0$, which is due to the
topological nature of the subgap states as discussed in
\secref{sec:shortNS}. $E_C$ is small compared with $E_M$ and $E_Z$ because the
Cooper pair tunneling, via the normal Andreev reflection, is higher-order
process: Note that the $E_M$ and $E_Z$-terms originates from a single electron
circulation around the ring.

\begin{figure}[t]
  \centering
  \includegraphics[width=8cm]{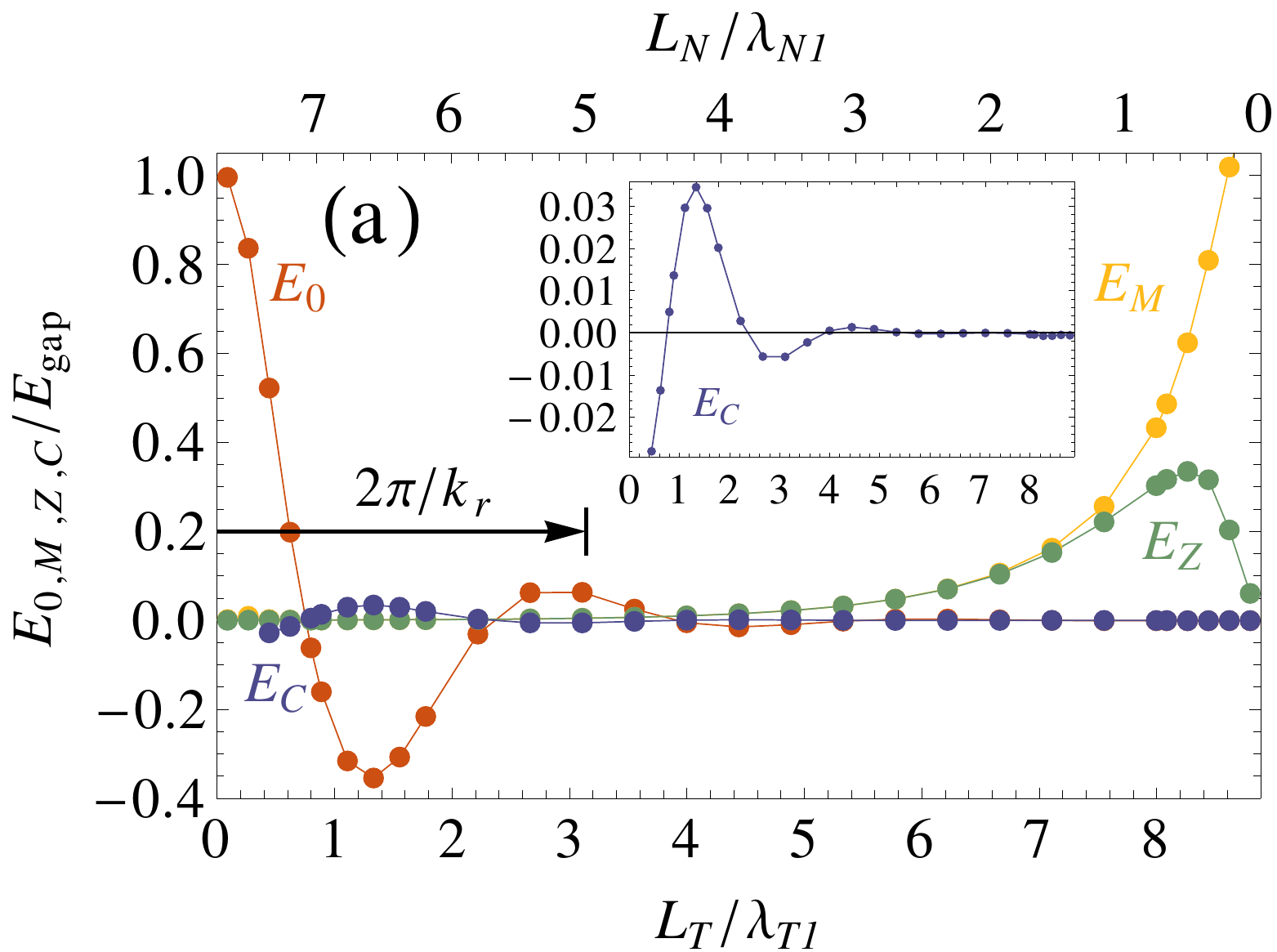}\\
  \includegraphics[width=8cm]{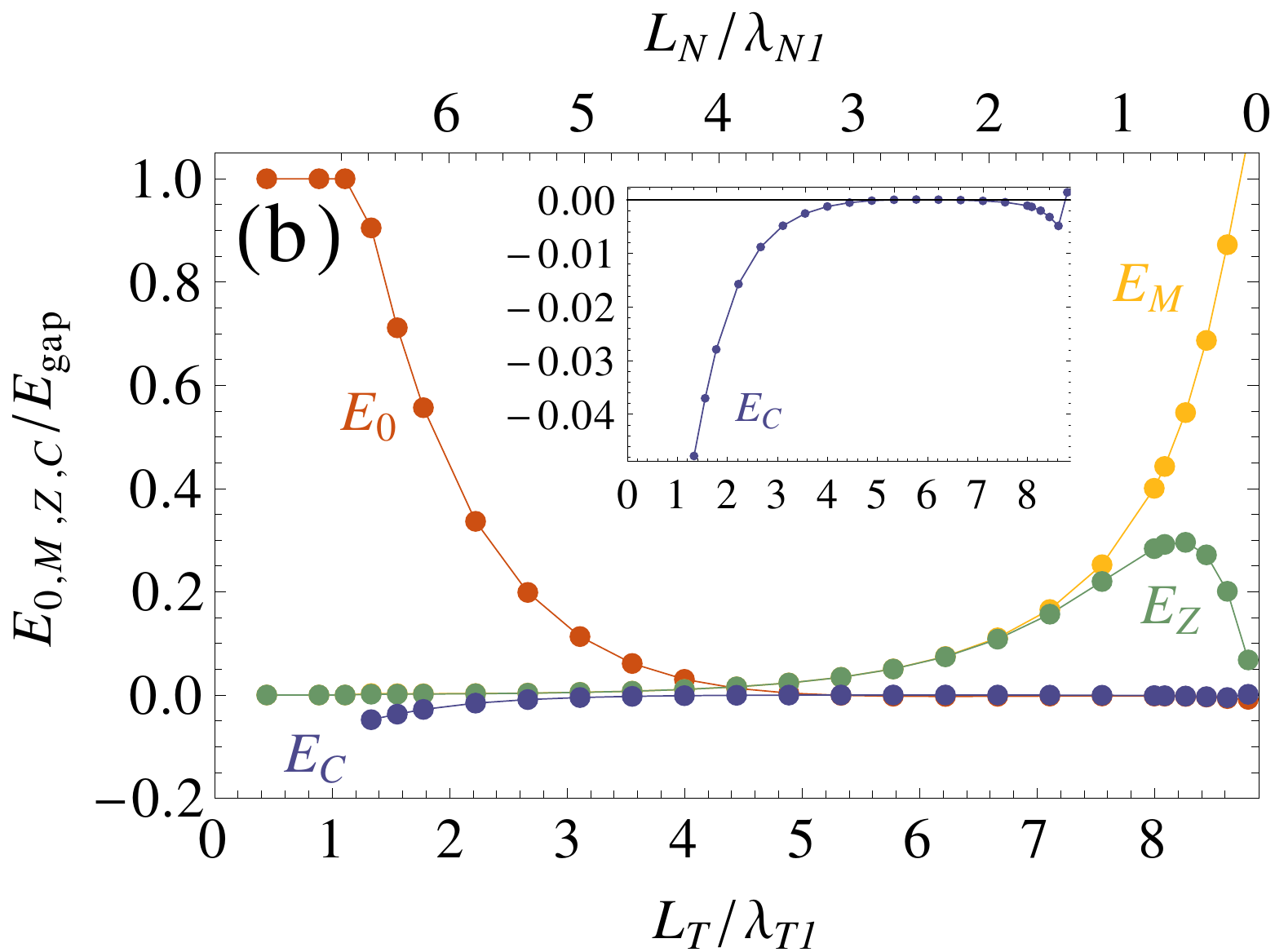}
  \caption{(Color online) The coefficients $E_0$, $E_C$, $E_M$, and $E_Z$ as
    functions of $L_N$ or $L_T=L-L_N$ with $L = L_N+L_T$ fixed. The inset shows
    the enlarged view of $E_C$. The parameters used are $\tilde\Delta = 3$,
    $\tilde\mu_N = -5$, and $\tilde\mu_T = 10$ (deep topological phase) [(a)]
    and 2.5 (weak topological phase) [(b)].}
  \label{fig:L:vst_wt}
\end{figure}

As pointed out in the previous section, \figref{fig:L:st}(a) shows
sign-changing oscillatory behaviors of the coefficients $E_0$ and $E_C$ when
the TS region is in the deep topological phase, $\tilde\mu_T(=5)>D(=2)$. The
period of the oscillation is $2\pi/k_r$ as expected from the sinusoidal
dependence in \eqnref{eq:E0:pt}.  In deeper topological phase ($\tilde\mu_T
\gg D$), the period becomes shorter [see \figref{fig:L:vst_wt}(a)], resulting
in more number of oscillations of $E_0$ and $E_c$ before they are
suppressed. Hence, the sign of the supercurrent can be controlled not only by
tuning the TS segment length $L_T$ with $k_r$ fixed but also by changing the
period $2\pi/k_r$ with $L_T$ fixed. The latter control can be done by tuning
the chemical potential $\tilde\mu_T$: Note that $k_r = \sqrt{\tilde\mu_T -
  D}$. If the TS region is in the weak topological phase, no oscillation is
observed [see \figref{fig:L:vst_wt}(b)] and the monotonic dependence of the
coefficients $E_0$ and $E_C$ on $L_T$ is observed.

Finally, we examine the length dependence of the phase shifts shown in
\figref{fig:L:st}(b). The phase $\gamma_Z$ is found to be zero, irrespective
of the segment length. It implies that the curvature of the ring does not
affect the transport due to the CAR process.
In contrast, the phase $\gamma_M$ is finite for all the length. In the case of
$L_N>\lambda_{N1}$ and $L_T>\lambda_{T1}$, where the perturbation is valid, the
phase is given by $\pi - 2\gamma$, the phase shift for the $\nu=1$ mode [see
\eqnref{eq:hm}] for $\gamma'\approx0$. For $L_N<\lambda_{N1}$, both $\nu=1,2$
modes are contributing so that the phase $\gamma_M$ becomes
length-dependent. The phase $\gamma_M$ becomes negligible only when the size of
the ring is sufficiently large: in this case $\tilde\Delta\gg1$ and
$\gamma\approx\pi/2$ [see \eqnref{eq:gamma}]. The phase shift $\gamma_M$ comes
from the finite curvature of the ring and the resultant phase shift of the
Majorana states. Our results show that this phase appears only in the
$E_M$-term, not in the $E_Z$ and, more importantly, $E_C$ terms.  Hence, the
existence of the finite phase shift $\gamma_M$ can be used for the evidence of
the Majorana fermions.

\subsection{Small Rings ($L_N\sim\lambda_{N1}$, $L_T\sim\lambda_{T1}$)}
\label{sec:smallRings}

\begin{figure}[t]
  \centering
  \includegraphics[width=7cm]{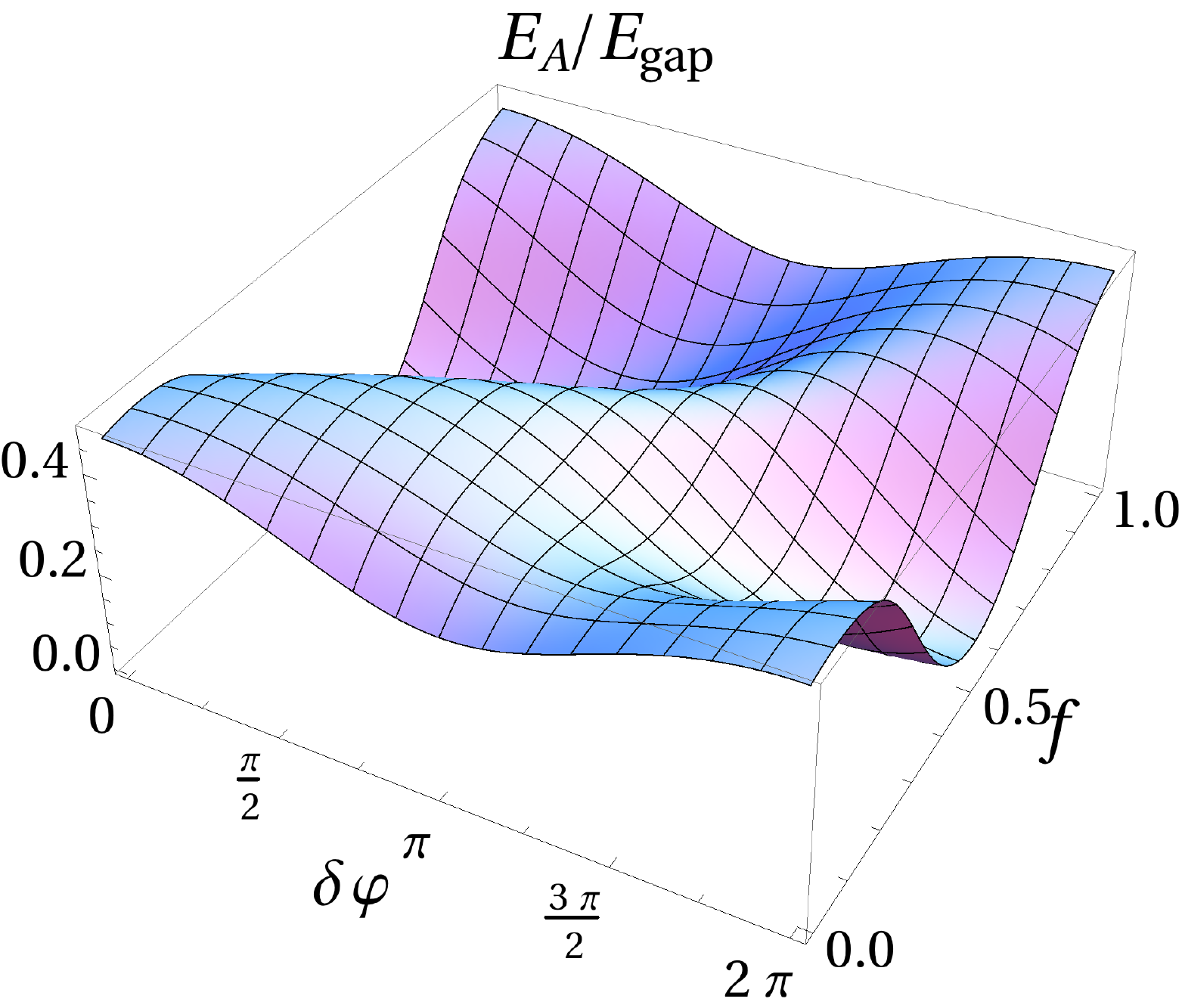}
  \caption{(Color online) Subgap energy for the state $\ket{1}$ as a function
    of $\delta\varphi$ and $f$ for the short non-topological region and short
    topological region: $L_N/\lambda_{N1} \approx 1.84$ and $L_T/\lambda_{T1}
    \approx 1.81$.  Here we have used $\tilde\Delta = 3$, $\tilde\mu_N = -2$,
    $\tilde\mu_T = 1.3$, and $E_\mathrm{gap} \approx 25\mu\mathrm{eV}$.}
  \label{fig:subgapE:LN500}
\end{figure}

\begin{figure}[t]
  \centering
  \includegraphics[width=8cm]{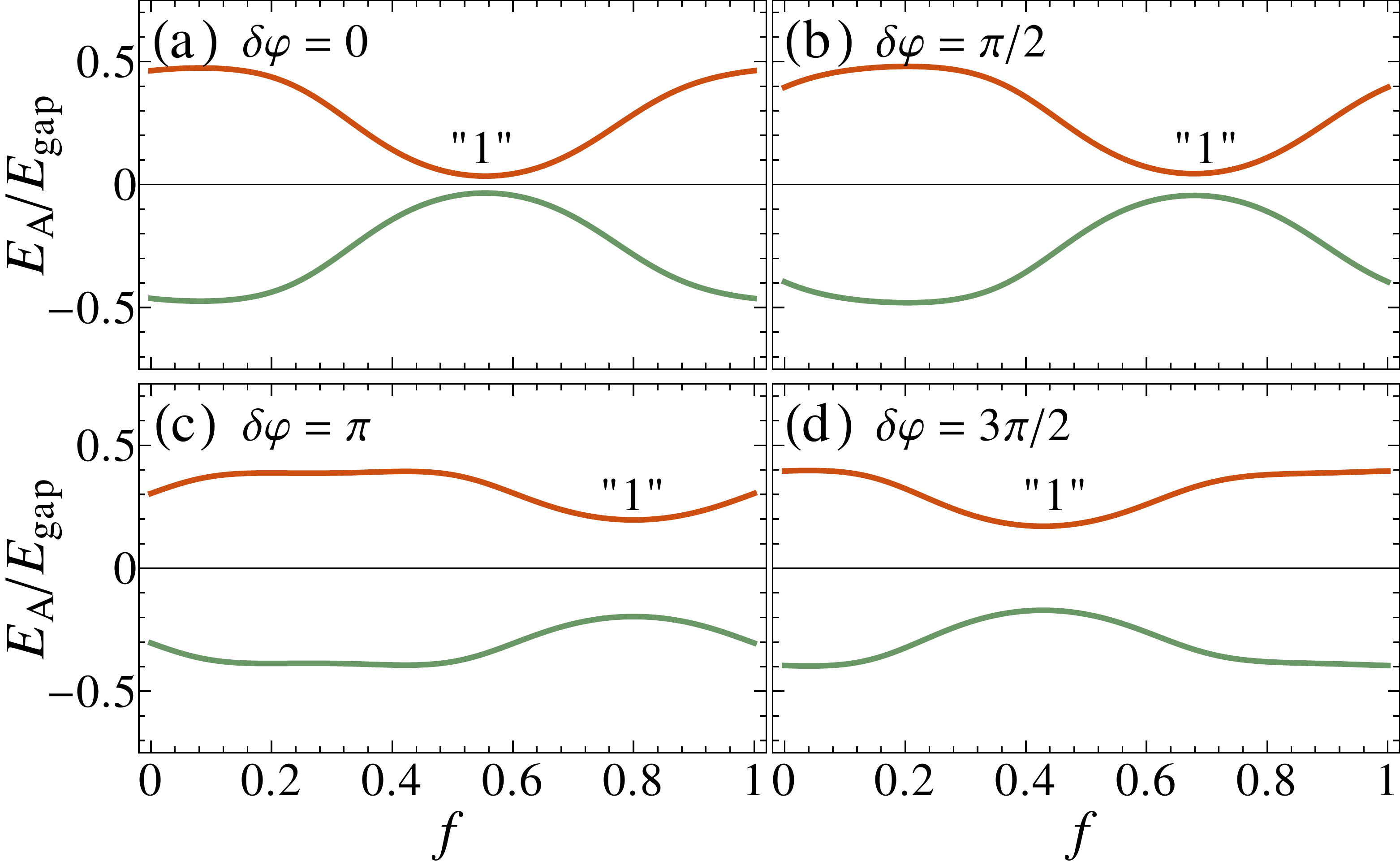}
  \caption{(Color online) Subgap energies as a function of $f$ for fixed values
    of $\delta\varphi$ as annotated. The energy which corresponds to the state
    $\ket{1}$ is marked by ``1''. We have used the same values for parameters
    as used in \figref{fig:subgapE:LN500}.}
  \label{fig:subgapE:LN500:dphi}
\end{figure}

Up to now, we have considered the cases in which the ring is large enough that
only one of CAR and NAR processes is operative. However, if the ring is small
or the localization length of the Majorana fermion is comparable to the
circumference of the ring, both processes can coexist. The general form of the
subgap energy is then given by
\begin{align}
  \begin{split}
    E_A
    & \approx
    E_0 + E_C \left[\cos \delta\varphi + \cos(4\pi f - \delta\varphi)\right]
    \\
    & \quad\mbox{}
    + E_M \cos (2\pi f + \gamma_M)
    \\
    & \quad\mbox{}
    + E_Z \cos(\delta\varphi - 2\pi f + \gamma_Z).
  \end{split}
\end{align}
In \figsref{fig:subgapE:LN500} and \ref{fig:subgapE:LN500:dphi}, we present the
subgap energy in the case where both $L_N$ and $L_T$ are comparable to the
Majorana fermion size, $L_N/\lambda_{N1} \sim L_T/\lambda_{T1} \sim 2$. In this
case we obtain
\begin{align*}
  \begin{split}
    \frac{E_0}{E_\mathrm{gap}}
    & \approx 0.30,
    \quad
    \frac{E_C}{E_\mathrm{gap}}
    \approx -0.02,
    \\
    \frac{E_M}{E_\mathrm{gap}}
    & \approx 0.13,
    \quad
    \frac{E_Z}{E_\mathrm{gap}}
    \approx 0.12,
    \quad
    \gamma_M
    \approx -0.23,
    \quad
    \gamma_Z
    \approx 0.
  \end{split}
\end{align*}
While the $E_C$-term is still small due to its nature of high-order processes,
the other terms are comparable. Since the $E_C$-term is negligible, the
supercurrent through the ring is entirely due to the CAR process.
The most intriguing point here is that the $f=1$ periodicity is protected even
if there is fermion parity breaking. \Figref{fig:subgapE:LN500:dphi} shows that
there is no crossing between the state $\ket0$ and $\ket1$. The constant
$E_0$-term, larger than the other terms, makes a big energy separation between
$\ket0$ and $\ket1$ states so that they are not coupled even if there is
fermion parity breaking. This kind of protection of $f=1$ periodicity was
also noticed in Ref.~\onlinecite{Pientka13a}. In our system, this
protection not only guarantees to observe the $f=1$ periodicity but also
provides us with a way to observe the perfect CAR process without other
deterioration.

\section{Conclusion}
\label{sec:conclusion}

We have considered a system of TS-NS double junctions in ring geometry to
investigate its supercurrent characteristics associated with the underlying
topological properties and Majorana subgap states localized at the junctions.
The system allows us to study on an equal footing TS-NS-TS and NS-TS-NS double
junction, which turn out to have topologically distinct supercurrent
characteristics. In this setup, the relative phases across the junctions are
controlled by the Aharonov-Bohm phase from the threading magnetic flux as well
as the phase difference between bulk the superconductors that induce the
$p$-wave superconductivity in the nanowire.

We have found that TS-NS-TS and NS-TS-NS double junction, seemingly counterpart
of each other, have substantially different supercurrent characteristics due to
the topological properties of their subgap states.
In our ring geometry containing both types of double junction, the supercurrent
characteristics depend strongly on the ratios of the wire segment lengths and
the localization lengths of the Majorana states.
For short (compared with the localization lengths of the Majorana states) NS
and long TS segment (\Secref{sec:shortNS}), the supercurrent originates solely
from the crossed Andreev reflection, exhibiting an unusual dependence on the
magnetic flux.
For short TS and long NS segment (\Secref{sec:shortTS}), on the contrary, the
normal Andreev reflection (NAR) determines the supercurrent, whose sign can be
oscillatory with the TS segment length. The difference in the supercurrent
features of the two extreme cases is explained in terms of topological
properties in the subgap states (\Secref{sec:topology}).
The representative characteristics in the above two extreme cases compete with
each other and show rich effects, which we study by varying the lengths of NS
and TS segment (\Secref{sec:length} and \ref{sec:smallRings}).

\begin{acknowledgments}
  This work was supported by the National Research Foundation of Korea (NRF)
  grants funded by the Korea MEST (Nos. 2011-0030790, 2011-0012494, and
  2010-0025880).
\end{acknowledgments}

\appendix

\section{Perturbative Approach}
\label{sec:perturbation}

In order to obtain an analytical expression for the subgap energy $E_A$ as a
function of $\delta\varphi$ and $f$, we take a perturbative approach,
performing a similar calculation used in Refs.~\onlinecite{Jiang11a} and
\onlinecite{Pientka13a}. First we neglect the interaction between two Majorana
states, each of which is localized at the interface, through the TS and NS
regions. In our ring geometry, it is done by applying the boundary conditions
only at one of the boundaries, see \eqnref{eq:bc}. Suppose that $\Psi_a(x)$ and
$\Psi_b(x)$ are the unperturbed Majorana wave functions localized at $x = x_a$
and $x = b$, respectively. Then, $\Psi_b(x)$ is the zero-energy eigenstate of
the Hamiltonian, \eqnref{eq:Hbdg} in the region $0 < x < L$ including the
boundary $x = x_b$, but not $x = x_a$. Therefore, $\Psi_b(x)$ satisfies the
boundary condition at $x = b$ only. Similarly, $\Psi_a(x)$ is defined in the
region $L_N < x < L+L_N$ and satisfies the boundary conditions at $x=x_a$
only. The wave functions are then given by linear combinations of the
zero-energy eigenstates in \eqnref{eq:wavefunction:zeroenergy}, whose
coefficients are determined via the boundary conditions,
\eqnref{eq:bc}. Explicitly, the normalized wave functions for $\Psi_i(x)$
($i=a,b$) are
\begin{align}
  \label{eq:Majoranawavefunction}
  \Psi_i(x)
  =
  \begin{cases}
    \Psi_{iN}(x),
    & x_a < x < x_b
    \\
    \Psi_{iT}(x),
    & x_b < x < L
  \end{cases}
\end{align}
with
\begin{subequations}
  \begin{align}
    \Psi_{aN}(x)
    & = \sum_\nu \frac{c_{aN\nu}}{\sqrt{N_a}} \chi^N_{+,\nu}(x)
    \\
    \Psi_{aT}(x)
    & =
    \sum_\nu \frac{c_{aT\nu}}{\sqrt{N_a}}
    e^{(i\nu k_r-1/\lambda_{N\nu})L} \chi^T_{-,\nu}(x)
    \\
    \Psi_{bN}(x)
    & =
    \sum_\nu \frac{c_{bN\nu}}{\sqrt{N_b}} e^{-L_N/\lambda_{N\nu}} \chi^N_{-,\nu}(x)
    \\
    \Psi_{bT}(x)
    & =
    \sum_\nu \frac{c_{bT\nu}}{\sqrt{N_b}}
    e^{-(i\nu k_r-1/\lambda_{N\nu})L_N} \chi^T_{+,\nu}(x)\ .
  \end{align}
\end{subequations}
Note here that the additional exponential factors have been inserted to make
the coefficients of order one at the localization center. The coefficients are
\begin{subequations}
  \label{eq:ca}
  \begin{align}
    c_{aN1}
    & = \frac{\sin(\gamma - 2\pi(f+1/2) + \delta\varphi/2)}{\sin\gamma}
    \\
    c_{aN2}
    & = e^{i\gamma} \frac{\sin(2\pi(f+1/2)-\delta\varphi/2)}{\sin\gamma}
    \\
    c_{aT1}
    & = \frac12 \left(1 + \sqrt{\frac{D-\tilde\mu_N}{D-\tilde\mu_T}}\right)
    \\
    c_{aT2}
    & = \frac12 \left(1 - \sqrt{\frac{D-\tilde\mu_N}{D-\tilde\mu_T}}\right)
  \end{align}
\end{subequations}
and
\begin{subequations}
  \label{eq:cb}
  \begin{align}
    c_{bN1}
    & = \frac{\sin(\gamma - \delta\varphi/2)}{\sin\gamma}
    \\
    c_{bN2}
    & = e^{-i\gamma} \frac{\sin(\delta\varphi/2)}{\sin\gamma}
    \\
    c_{bT1}
    & = \frac12 \left(1 + \sqrt{\frac{D-\tilde\mu_N}{D-\tilde\mu_T}}\right)
    \\
    c_{bT2}
    & = \frac12 \left(1 - \sqrt{\frac{D-\tilde\mu_N}{D-\tilde\mu_T}}\right)
  \end{align}
\end{subequations}
The normalization constants are given by
\begin{align}
  \begin{split}
    N_i
    & =
    \frac{1}{R}
    \sum_\nu
    \left[
      \frac{|c_{iN\nu}|^2}{\lambda_{N\nu}^{-1}}
      +
      2\cos\gamma \frac{|c_{iN\nu} c_{iN\bar\nu}|}%
      {\lambda_{N1}^{-1}+\lambda_{N2}^{-1}}
    \right.
    \\
    & \qquad\quad\left.\mbox{}
      +
      \frac{|c_{iT\nu}|^2}{\lambda_{T\nu}^{-1}}
      +
      \frac{2 c_{iT\nu}^* c_{iT\bar\nu}}%
      {2(-1)^\nu i k_r + \lambda_{T1}^{-1}+\lambda_{T2}^{-1}}
    \right].
  \end{split}
\end{align}
Here the normalization constants are obtained up to the leading order in the
small factor $e^{-L/\lambda_{\ell\nu}}$, which is consistent with our
perturbation.  Note that the Rashba phase appear explicitly in the coefficients
$c_{aN\nu}$ in the form of $f+1/2$, as discussed in \secref{sec:bulk}.

The effective Hamiltonian projected to the Majorana subspace is then
represented as
\begin{align}
  H_\mathrm{M}
  =
  \begin{bmatrix}
    \Braket{\Psi_a|H_\mathrm{eff}^\mathrm{BdG}|\Psi_a}
    & \Braket{\Psi_a|H_\mathrm{eff}^\mathrm{BdG}|\Psi_b}
    \\
    \Braket{\Psi_b|H_\mathrm{eff}^\mathrm{BdG}|\Psi_a}
    & \Braket{\Psi_b|H_\mathrm{eff}^\mathrm{BdG}|\Psi_b}
  \end{bmatrix}
\end{align}
Since $\Psi_i(x)$ are not the eigenstates of the full Hamiltonian $H_{\rm
  eff}^\mathrm{BdG}$, the diagonal terms does not vanish. However, we ignore
them since they are proportional to the square of the exponential factor
$e^{-L/\lambda_{\ell\nu}}$ and much smaller than the off-diagonal terms. The
formal expression for the off-diagonal terms are
\begin{subequations}
  \begin{align}
    \nonumber
    H_{M,ab}
    & =
    i
    \left\{
      \Psi_a^\dag(x_a) (v_\phi \Psi_b(x_a^-) - v_\phi \Psi_b(x_a^+))
    \right.
    \\
    & \qquad\left.\mbox{}
      + [v_\phi \Psi_a(x_a)]^\dag (\Psi_b(x_a^-) - \Psi_b(x_a^+))
    \right\}
    \\
    \nonumber
    H_{M,ba}
    & =
    - i
    \left\{
      \Psi_b^\dag(x_b) (v_\phi \Psi_a(x_b^+) - v_\phi \Psi_a(x_b^-))
    \right.
    \\
    & \qquad\left.\mbox{}
      + [v_\phi \Psi_b(x_b)]^\dag (\Psi_a(x_b^+) - \Psi_a(x_b^-))
    \right\}\ .
  \end{align}
\end{subequations}
Since $\Psi_a(x)$ and $\Psi_b(x)$ are not orthogonal to each other, the
effective Hamiltonian is not necessarily hermitian, $H_{M,ab} \ne
H_{M,ba}^*$. The subgap energy is then obtained as
\begin{align}
  E_A = \pm \sqrt{H_{M,ab} H_{M,ba}}.
\end{align}
Explicit and tedious calculations lead to
\begin{subequations}
  \begin{align}
    H_{M,ab}
    & =
    + i E_R \frac{e^{-i\gamma}}{\sqrt{N_aN_b}}
    \sum_{\ell\nu} e^{-L_\ell/\lambda_{\ell\nu}} h^-_{\ell\nu}
    \\
    H_{M,ba}
    & =
    - i E_R \frac{e^{+i\gamma}}{\sqrt{N_aN_b}}
    \sum_{\ell\nu} e^{-L_\ell/\lambda_{\ell\nu}} h^+_{\ell\nu}
  \end{align}
\end{subequations}
with
\begin{align}
  \label{eq:hm}
  h^\pm_{N\nu}
  =
  \frac{(-1)^\nu \epsilon_1}{\sin\gamma}
  \left[
    \cos(2\pi f {+} \gamma' {-} \zeta_\nu)
    - \cos(\delta\varphi{-}2\pi f{\pm}\gamma')
  \right]
\end{align}
with $\zeta_1 = 2\gamma$ and $\zeta_2 = 0$ and
\begin{align}
  h^\pm_{T\nu}
  =
  \epsilon_1 \sin(\gamma-\gamma')
  - (-1)^\nu (\epsilon_2 \sin\gamma - \epsilon_3 \cos\gamma)
\end{align}
for $0 < \tilde\mu_T < D$ and
\begin{align}
  \begin{split}
    \sum_\nu h^\pm_{T\nu}
    & =
    2 \epsilon_1 \sin(\gamma-\gamma') \cos k_r L_T
    \\
    & \quad\mbox{}
    +
    2 \left(\epsilon_2 \sin\gamma - \epsilon_3 \cos\gamma\right) \sin k_r L_T\ .
  \end{split}
\end{align}
for $D < \tilde\mu_T$. Here we have defined
\begin{subequations}
  \label{eq:h}
  \begin{align}
    \epsilon_1 & \equiv \sqrt{\tilde\Delta^2-4\tilde\mu_N}
    \\
    \epsilon_2
    & \equiv \frac{2D-\tilde\mu_N-\tilde\mu_T}{\sqrt{|D-\tilde\mu_T|}}
    \\
    \epsilon_3 & \equiv \frac{\sqrt{D-\tilde\mu_N}}{\sqrt{|D-\tilde\mu_T|}}
  \end{align}
\end{subequations}
and
\begin{align}
  \cos\gamma'
  & \equiv \sqrt{1 - 1/\epsilon_1^2},
  &
  \sin\gamma'
  & \equiv 1/\epsilon_1\ .
\end{align}

In the large curvature limit $(R\to\infty)$ where $\gamma\to\pi/2$ and
$\gamma'\to0$, the coefficients are simplified to
\begin{subequations}
  \begin{align}
    h^\pm_{N\nu}
    & =
    \epsilon_1 \left[\cos2\pi f - (-1)^\nu\cos(\delta\varphi - 2\pi f)\right]
    \\
    h^\pm_{T\nu}
    & = \epsilon_1 - (-1)^\nu \epsilon_2
    \quad (\tilde\mu_T < D)
    \\
    \sum_\nu h^\pm_{T\nu}
    & = 2 \epsilon_1 \cos k_r L_T + 2 \epsilon_2 \sin k_r L_T
    \quad (\tilde\mu_T > D)
  \end{align}
\end{subequations}
and
\begin{subequations}
  \begin{align}
    N_a
    & = N_0 + \frac{\sqrt{D}}{-\tilde\mu_N} \cos(4\pi f-\delta\varphi),
    \\
    N_b
    & = N_0 + \frac{\sqrt{D}}{-\tilde\mu_N} \cos\delta\varphi
  \end{align}
\end{subequations}
with
\begin{align}
  \label{eq:N}
  N_0
  \equiv
  \frac{\sqrt{D-\tilde\mu_N}}{-\tilde\mu_N}
  +
  \frac{\sqrt{D} + \sqrt{D-\tilde\mu_N}}{\tilde\mu_T}
  +
  \frac{\tilde\mu_T - \tilde\mu_N}{2\tilde\mu_T\sqrt{D}}
\end{align}
where $N_0$ is the value of the normalization constants $N_{a,b}$ averaged over
the phases.

\end{document}